\documentclass[aps,reprint,superscriptaddress,amsmath,amssymb]{revtex4-2}

\usepackage{xr-hyper}
\usepackage{graphicx}
\usepackage{listings}
\usepackage{longtable}
\usepackage{color,soul}
\usepackage{algorithm}
\usepackage[bookmarks,bookmarksnumbered,colorlinks,allcolors=blue]{hyperref}
\usepackage{bookmark}
\usepackage{makecell}
\usepackage{physics}
\usepackage{xcolor}
\usepackage{lineno}
\usepackage{xcolor}

\definecolor{dkgreen}{rgb}{0,0.6,0}
\definecolor{gray}{rgb}{0.5,0.5,0.5}
\definecolor{mauve}{rgb}{0.58,0,0.82}

\newcommand{\Phiext}{\ensuremath{\Phi_{\text{ext}}}}
\newcommand{\TixAlxNraw}{Ti\textsubscript{x}Al\textsubscript{1-x}N}
\newcommand{\TixAlxN}{Ti\textsubscript{x}Al\textsubscript{1-x}N\ }
\newcommand{\Teff}{T_{\text{eff}}}

\newcommand{\ATone}{\ensuremath{A_{\Phi}}}
\newcommand{\ATtwo}{\ensuremath{A_{\Phi}}}
\newcommand{\omegaref}{\ensuremath{\omega_{\textrm{ref}}}}

\newcommand*{\addFileDependency}[1]{
\typeout{(#1)}
\IfFileExists{#1}{}{\typeout{No file #1.}}
}\makeatother

\begin{document}

\title{\Large The effects of disorder in superconducting materials on qubit coherence}

\author{Ran Gao}
\thanks{These authors contributed equally to this work.}
\affiliation{Quantum Science Center of Guangdong-Hong Kong-Macao Greater Bay Area, Shenzhen 518045, China}
\affiliation{Z-Axis Quantum, Hangzhou, China}

\author{$^{\!\!,\;\textcolor{blue}\dagger}$ Feng Wu}
\thanks{These authors contributed equally to this work.}
\affiliation{Zhongguancun Laboratory, Beijing, China}

\author{Hantao Sun}
\thanks{These authors contributed equally to this work.}
\affiliation{China Telecom Quantum Information Technology Group Co., Ltd., Hefei 230031, China}

\author{Jianjun Chen}
\affiliation{Xinxiao Electronics Inc., Hangzhou 310018, China}

\author{Hao Deng}
\affiliation{International Center for Quantum Materials, Peking University, Beijing 100871, China}

\author{Xizheng Ma}
\affiliation{Quantum Science Center of Guangdong-Hong Kong-Macao Greater Bay Area, Shenzhen 518045, China}

\author{Xiaohe Miao}
\affiliation{Instrumentation and Service Center for Molecular and Physical Sciences and Research Center for Industries of the Future, Westlake University, Hangzhou 310030, China}

\author{Zhijun Song}
\affiliation{Shanghai E-Matterwave Sci \& Tech Co., Ltd., Shanghai 201100, China}

\author{Xin Wan}
\affiliation{Zhejiang Institute of Modern Physics and Zhejiang Key Laboratory of Micro-nano Quantum Chips and Quantum Control, Zhejiang University, Hangzhou 310027, China}

\author{Fei Wang}
\affiliation{Quantum Science Center of Guangdong-Hong Kong-Macao Greater Bay Area, Shenzhen 518045, China}
\affiliation{Z-Axis Quantum, Hangzhou, China}

\author{Tian Xia}
\affiliation{Huaxin Jushu Microelectronics Co., Ltd., Hangzhou 310052, China}

\author{Make Ying}
\affiliation{EXTEC Inc., Hangzhou 310024, China} 

\author{Chao Zhang}
\affiliation{Instrumentation and Service Center for Physical Sciences, Westlake University, Hangzhou 310024, Zhejiang, China}

\author{Yaoyun Shi}
\affiliation{Z-Axis Quantum, Hangzhou, China}

\author{Hui-Hai Zhao}
\affiliation{Zhongguancun Laboratory, Beijing, China}

\author{Chunqing Deng}
\email{gaoran@quantumsc.cn; \\ dengchunqing@quantumsc.cn}
\affiliation{Quantum Science Center of Guangdong-Hong Kong-Macao Greater Bay Area, Shenzhen 518045, China}
\affiliation{Z-Axis Quantum, Hangzhou, China}

\begin{abstract}

Introducing disorder in the superconducting materials has been considered promising to enhance the electromagnetic impedance and realize noise-resilient superconducting qubits. Despite a number of pioneering implementations, the understanding of the correlation between the material disorder and the qubit coherence is still developing. Here, we demonstrate a systematic characterization of fluxonium qubits with the superinductors made by spinodal titanium-aluminum-nitride with varied disorder. From qubit noise spectroscopy, the flux noise and the dielectric loss are extracted as a measure of the coherence properties. Our results reveal that the $1/f^\alpha$ flux noise dominates the qubit decoherence around the flux-frustration point, strongly correlated with the material disorder; while the dielectric loss are largely similar under a wide range of material properties. From the flux-noise amplitudes, the areal density ($\sigma$) of the phenomenological spin defects and material disorder are found to be approximately correlated by $\sigma \propto \rho_{xx}^3$, or effectively $(k_F l)^{-3}$. This work has provided new insights on the origin of decoherence channels beyond surface defects and within the superconductors, and could serve as a useful guideline for material design and optimization.

\end{abstract}

\maketitle
\bookmarksetup{startatroot}

\clearpage

\section*{Introduction}

\begin{figure*} [t]
      \centering
      \includegraphics[width=12.9cm]{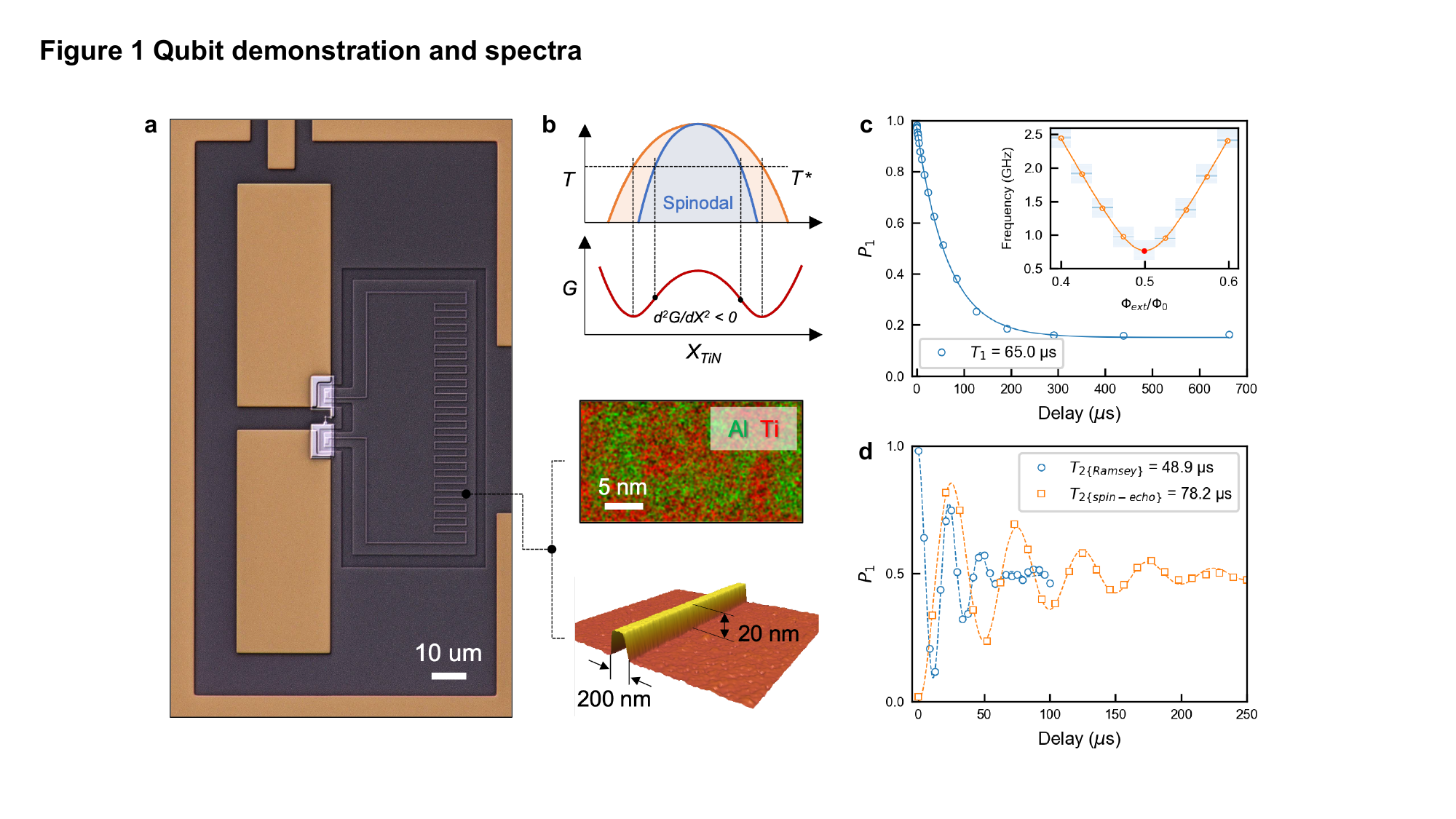}
      \caption{\textbf{Fluxonium qubit based on disordered spinodal superconductor} (a) The optical image of a fluxonium qubit with \TixAlxN wire as the superinductor. The image was taken using the optical-shadow mode where the topography of the surface features are mapped and exaggerated by utilizing incident lights from multiple directions (\textit{e.g.}, see the over-etched sapphire features and the contact leads of the inductor wire underneath the aluminum pads). The energy dispersive spectroscopy (EDS) scan on a cross-sectional region of the \TixAlxN films and the atomic force microscope (AFM) measurement on the inductor wires are provided. Here, the measured device consists a 200 nm-wide inductor wire, which is the narrowest we have used in this study. The AFM studies revealed a clean surface and well-defined edges, and similar morphology were found for wires with alternative dimensions. (b) A schematic illustration of the spinodal decomposition in the \TixAlxN. The temperature $T$ and \textit{Gibbs} free energy $G$ are plotted against the concentration of the TiN phase. $T^*$ marks the temperature at which the \textit{Gibbs} free energy curve is plotted. The spinodal decomposition takes place at compositions where the second derivative of the \textit{Gibbs} free energy is negative. (c-d) Energy relaxation $T_1$ and dephasing $T_2$ decays of a best sample. The inset is the qubit spectrum of this particular sample and the red dot labels the minimum qubit frequency at $\sim$768 MHz. }
      \label{fig:TANNQFig1} 
\end{figure*}

Leveraging disorder in superconducting materials has emerged as an important approach to manipulate the circuit properties for quantum information processing with superconducting qubits~\cite{Kjaergaard2020, Sacepe2020}. Demonstrated circuit elements such as superinductors~\cite{Grunhaupt2019, Hazard2019}, coherence quantum phase-slip junctions~\cite{Astafiev2012, deGraaf2018, Rieger2023}, and compact resonators~\cite{Zmuidzinas2012} are at the essence of engineering qubits with intrinsic noise protection and the integration of a large-scale quantum processor~\cite{Manucharyan2009, Pechenezhskiy2020, Kalashnikov2020, Gyenis2021, Brooks2013, Le2019}. Key to such applications, albeit still developing, is the understanding of the interplay between material disorder and quantum coherence. In particular, the evolution of the coherence properties of the superconducting materials, as with the increase in material disorder, is of scientific and practical significance in implementing the corresponding quantum circuits.

Despite the extensive research on the impact of material defects on qubit coherence~\cite{Oliver2013, Murray2021, Siddiqi2021}, the learnings are yet to be applied directly on guiding the design of the disordered systems. This is partly due to the fact that the majority of the studies are typically in the limit where the penetration depth ($\lambda$) is comparable or smaller than the thickness ($t$) and width ($w$) of the superconductors. In other words, the electromagnetic fields are concentrated at the material surfaces, edges, or the metal-insulator boundaries. As a consequence of the limited volume, the exact material properties at these regimes are largely depending on the fabrication history, challenging to be quantitatively characterized, and are often, if not always, postulated~\cite{Paladino2014, Muller2019, DeGraaf2022}. At the opposite end, a disordered superconducting wire features a large kinetic inductance is in the $\lambda > w \gg t$ regime where the current distribution within the material is essentially homogeneous~\cite{Clem2013}. As such, the bulk material properties that can be reliably characterized could serve as the figure of merit of a certain material, and thus be used for quantitative analysis.

Driven by the above, we introduce fluxonium qubits with superinductors made by spinodal titanium-aluminum-nitride (\TixAlxNraw) thin films~\cite{Gao2022a}, and focus on qubit coherence as a function of the material disorder (quantified by the longitudinal resistivity $\rho_{xx}$). For different set of qubits, the disorder of the titanium-aluminum-nitride is tuned by a combined variation of the film thickness, chemical compositions, and annealing conditions. For each qubit, we utilize the qubit as a spectrometer for noise~\cite{Bylander2011} and quantitatively extract the dielectric loss and $1/f^\alpha$ flux noise amplitudes that are dominant in decoherence at different ends of the qubit spectrum~\cite{Sun2023}. Our findings reveal that the dielectric loss tangent ($\tan\delta_C$) of the qubits takes a typical value around $\sim 1.5\mbox{-}4 \times 10^{-6}$ in the weak-disorder limit, and likely increases as the disorder is enhanced beyond $\rho_{xx} > 10^{-3}$ $\Omega$-cm. On the other hand, the $1/f^\alpha$ flux noise, with $\alpha$ ranging from $\sim$0.8 to $\sim$1.2, is found to be the dominant source of qubit decoherence, mainly determined by the material disorder. More intriguingly, the areal density of the phenomenological spin defects $\sigma$ deduced from the flux noise amplitudes is found to be proportional to $\sim \rho_{xx}^3$. We also discuss the possible origins and mechanisms of such correlation.

\section*{Results}

\subsection*{Device structure and qubit measurement}

The qubit design is inherited from our previous studies except that the inductive element is replaced by a long wire of superconducting \TixAlxN (\autoref{fig:TANNQFig1}a, more details in Supplementary Fig.~1-2) \cite{Bao2022, Gao2022a}. Briefly, the insulating \TixAlxN thin films, with varied chemical compositions and thicknesses, were first patterned and annealed to introduce the spinodal phase segregation and superconductivity (\autoref{fig:TANNQFig1}a-b). The annealing was followed by the deposition and patterning of TiN films as the backbone of the superconducting circuits \cite{Gao2022, Deng2023}. Eventually, the superinductor wires were patterned and galvanically connected to the rest of the circuits and the Josephson junction using the conventional shadow-evaporation techniques. The geometries of the inductor wires were carefully characterized and measured, revealing a clean qubit topography. Each qubit is capacitively coupled to a resonator for dispersive readout, and on-chip charge and flux lines are used for qubit excitation and frequency modulation. The cryogenic characterizations were performed in a dilution refrigerator at a base temperature below 10 mK (see Methods and Supplementary Information for details). 

The spectra of the qubits were first obtained as a function of the external flux $\Phiext$. For each qubit, the corresponding charging energy $E_C$, inductive energy $E_L$, and Josephson energy $E_J$ were obtained by fitting the spectrum to the fluxonium Hamiltonian $\hat{H}= 4 E_C \hat{n}^{2}+ (E_L/2) \left(\hat{\varphi} + \Phiext / \varphi_0 \right )^{2}-E_J \cos\hat{\varphi}$. The operational frequencies at the flux-frustration spot (\textit{i.e.}, $\Phiext=\Phi_0/2$) of all the tested qubits are around several hundred megahertz, and a typical spectrum around the flux-frustration spot is provided (see \autoref{fig:TANNQFig1}c inset, Supplementary Fig.~4, and Supplementary Table~1).

The coherence measurement was performed using a flux-pulse protocol. Namely, after parking the qubit at the flux-frustration spot using a static DC bias, a flux pulse is applied to shift the qubit to the targeting operational points. At each operational point, the energy and free induction decay curves were measured and fitted to extract $T_1$ and $T_2$ for further decoherence analysis. Here, coherence times at the flux-frustration spot of a best device are provided (\autoref{fig:TANNQFig1}c-d). We note that the measured $T_1 \sim 65~\mu s$ and $T_{2e} \sim 78.2~\mu s$ are the highest ever reported for fluxonium made with disordered materials. Meanwhile, the coherence times are found strongly depending on the material properties. Given the fact that the qubit coherence times are also highly sensitive to the qubit parameters, making it less ideal as an absolute measure of the qubit coherence, we proceed to develop an analysis protocol to better understand the decoherence properties and to quantitatively compare among samples.

\subsection*{Coherence property analysis\label{subsec:analysis}}

\begin{figure}[h]
    \centering
    \includegraphics[width=8.6cm]{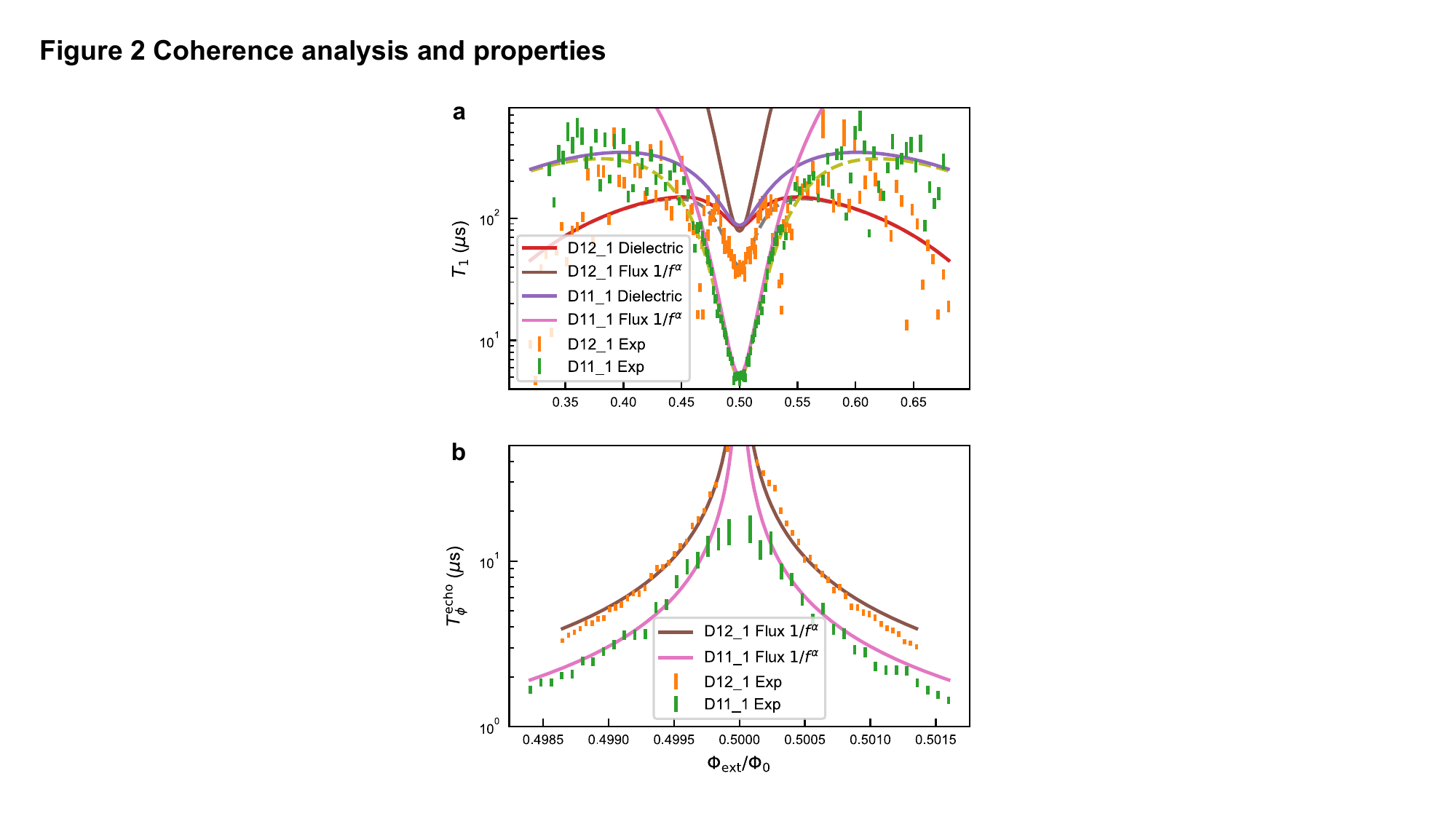}
    \caption{\textbf{The extraction of coherence properties} (a) The qubit relaxation time $T_1$ and (b) the qubit pure dephasing time from spin-echo measurements $T_{\phi}^{\text{echo}}$ versus the qubit external flux $\Phiext$ of two typical fluxonium qubits D11\_1 and D12\_1. The two qubits differ in the properties of the \TixAlxN wires and the qubit parameters (due to fabrication fluctuation, see Supplementary Table 1 for details). The markers and error bars represent the values and associated uncertainties of $T_1$ and $T_{\phi}$, which were derived by fitting experimental $P(t)$ curves using the least squares method. The solid lines are fitted values from a specific loss channel. The dashed lines in (a) are simulated $T_1$ which are close to the experiment data points. The flux noise dominates the $T_1$ when $\Phiext\approx 0.5 \Phi_0$ while the dielectric loss dominates when $\Phiext < 0.35 \Phi_0$ for these devices. 
    \label{fig:t1_t2_model}}
\end{figure}

The decoherence of the qubit can be characterized by the relaxation time $T_1$, the dephasing time $T_2$, and the pure dephasing time $T_{\phi}=(T_2-2/T_1)^{-1}$ or the corresponding rates $\Gamma_i=1/T_i(i=1,2,\phi)$. A notable feature is the suppressed $T_1$ around the flux-frustration point (\autoref{fig:t1_t2_model}a). It is found that the decoherence data can be reasonably well characterized by two major sources, the dielectric loss and the $1/f^\alpha$ flux noise~\cite{Sun2023} in the following form:
\begin{align}
    &\Gamma_1^{\text{diel}} = \frac{\hbar \omega_{01}^{2}}{4 E_C}  \abs{\mel{0}{\hat{\varphi}}{1}}^2 \tan\delta_C \coth\left( \frac{\hbar \omega_{01}}{2k_B\Teff}\right)\ , \label{eq:t1_dielectric}  \\
    &\Gamma_1^{\text{flux}} = \frac{2\pi E_L^2}{\hbar^2 \varphi_0^2}\abs{\mel{0}{\hat{\varphi}}{1}}^2 \frac{\ATone^2}{\omegaref(\omega_{01}/\omegaref)^\alpha} \left(1+ \exp\left(-\frac{\hbar\omega_{01}}{k_B\Teff}\right) \right)\ , \label{eq:t1_flux}
\end{align}
where $\Teff$ is the effective temperature, and $\tan\delta_C$, $\ATone$, $\alpha$ are the dielectric loss tangent, $1/f^\alpha$ flux noise amplitude, and the flux noise exponent respectively. $\omega_{\textrm{ref}}$ is the reference frequency selected to normalize the unit of the double-sided noise-power spectral spectral density $S_{\Phi}(\omega)=\frac{1}{2\pi}\int dt\left\langle\delta \Phi(0)\delta \Phi(t)\right\rangle e^{-i\omega t}= A_{\Phi}^2/\omegaref/(\omega/\omegaref)^{\alpha}$~\cite{Bylander2011}. Note that we can always scale $A_{\Phi}$ and $\omegaref$ together without changing $S_{\Phi}(\omega)$, and we keep $\omegaref/2\pi=1\textrm{Hz}$ in this study. The dephasing times measured from spin-echo measurements have a typical first-order coupling profile in the flux type qubits~\cite{Bylander2011}:

\begin{align}
    \Gamma_{\phi}^{\text{flux}} &= \left(\left(\frac{\partial \omega_{01}}{\partial \Phiext}\right)^2 \ATtwo^2 c(\alpha)\omegaref^{\alpha-1}\right)^{1/(1+\alpha)}, \label{eq:tphi_flux1f} \\
    c(\alpha) &= t^{1-\alpha}\int_0^{+\infty}d\omega\frac{1}{\omega^{\alpha}}\sin^2\left(\frac{\omega t}{4}\right)
    \textrm{sinc}^2\left(\frac{\omega t}{4}\right),
    \label{eq:tphi_alpha_coef}
\end{align}
where $c(\alpha)$ is the convolution between the noise spectrum and the filter function for the spin-echo experiments. 

A consistent data processing and fitting procedure was applied to all dataset, and by fitting the data with the above models we can extract $\tan\delta_C$, $\ATone$, and $\alpha$ (see Supplementary Information for details). To better illustrate the applicability of the analysis model, a comparison between the measured and the fitted data is provided (\autoref{fig:t1_t2_model}). As discussed above, the noise level is found to be dependent on the properties of the \TixAlxN, and two cases with distinct disorder are plotted together. Additional exemplary devices and sources of errors are further elaborated in the Supplementary Information (Supplementary Fig.~5-Fig.~7). 

It is worth noting that the energy relaxation around the flux-frustration point can be either interpreted by the flux noise~\cite{Yan2016, Quintana2017, Sun2023} or the inductive losses~\cite{Hazard2019} (Supplementary Fig.~6). Since the flux noise model inherently explains the dephasing behaviors that the variables to model our complicated systems can be minimized, we have chosen to use flux noise for this study. Meanwhile, it is acknowledged that if either the flux noise exponent or the inductive loss tangent is frequency dependent, these two models are essentially indistinguishable. Other dephasing mechanisms such as coherent quantum phase slips is unlikely as the $T_2$ is peaked other than suppressed at the flux-frustration spot~\cite{Manucharyan2012}. This is consistent with the fact that the inductor wires are quasi-2D with sufficient thickness and are far from the superconductor-to-insulator transition.

\subsection*{Tuning the material disorder}

\begin{figure*} [t]
      \centering
      \includegraphics[width=11.6cm]{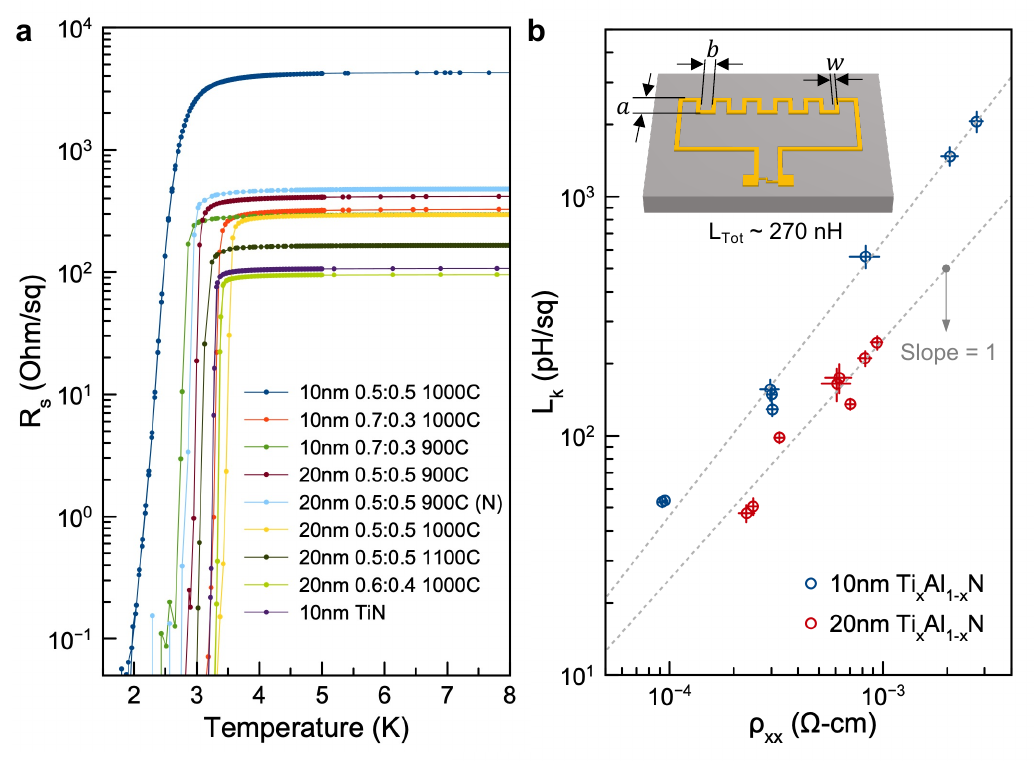}
      \caption{\textbf{Tuning the material disorder} (a) The temperature-dependent sheet resistance measured for \TixAlxN at a set of film thicknesses, chemical compositions, and annealing conditions. Data are labeled accordingly with the film thicknesses, nominal Ti:Al ratios from the deposition rates, and the annealing temperatures. Note that the annealing is all performed in the argon environment for 30 minutes except the one in the nitrogen environment labeled with (N). The 10 nm-thick TiN films was not annealed and was directly patterned for qubit devices. (b) The extracted $L_k$ from the qubit spectroscopy measurement as a function of the material disorder $\rho_{xx}$. The 20 nm-thick films yielded a good reciprocal relationship between $L_k$ and $\rho_{xx}$. The deviated slope in the 10 nm-thick films is due to the reduction of the superconductivity gap in thinner and more disordered films. The inset is an illustration of the controlled geometric parameters to achieve the target total inductance ($L_{tot}$) with different aspect ratio $p/w$, where $p$ and $w$ is the perimeter and width of the inductor wire, respectively. The total footprint of the loop is fixed while the three parameters $a$, $b$, and $w$ are adjusted. Besides the $p/w$ variation across wafers with different $L_k$, qubits with a fixed $p/w$ but different perimeters and widths were also measured on the same device for statistical consideration and as sanity checks of the flux-noise model (see more details in the Supplementary Information).} 
      \label{fig:TANNQFig3} 
\end{figure*}

With an established analysis protocol for the qubits, we then systematically vary the properties of the superinductor wires, and correlate those with the qubit coherence. Here, a representative set of \TixAlxN films were used in this work, and the temperature-dependent sheet resistance $R_s$ near the superconducting transition temperatures are given (\autoref{fig:TANNQFig3}a, details see Methods). We use the normal longitudinal resistivity ($\rho_{xx}=R_st$) at 10~K as a measure of the film disorder. The film thicknesses were measured using AFM, and are larger than the \textit{Ginzburg–Landau} superconducting coherence lengths ($\xi_{GL} \sim 7$ nm) of \TixAlxN~\cite{Gao2022a}. It is worth noting that we take disorder as the sole indicator of the material properties, regardless of the different experimental techniques applied to tune the microstructures. The rationalization and experimental grounding of this approach is detailed in the Supplementary Information. 

A span of two orders of magnitude in disorder is realized using the above sample set, and the correlation between the level of material disorder $\rho_{xx}$ and the kinetic inductance $L_k$ is illustrated (\autoref{fig:TANNQFig3}b). Here, the kinetic inductance is directly extracted and averaged from the corresponding qubit spectra using $L_k=(w/p) (\Phi_0/2\pi)^2/E_L$ (\autoref{fig:TANNQFig3}b inset and Supplementary Table 1). The 20 nm-thick films revealed an expected linear relationship between the kinetic inductance and the material disorder as $L_k = \hbar \rho_{xx} / \pi \Delta_0 t$, where $\Delta_0$ is the superconductivity gap at zero temperature. The slight deviation from such relationship in 10 nm-thick films is due to the reduced superconducting transition temperatures as a result of the increased disorder~\cite{Finkelstein1994, Carbillet2020, Sacepe2008, Sacepe2020}. 

For better investigation on the microscopic properties of the materials, as it will be discussed later, the celebrated \textit{Ioffe-Regel} parameter $k_F l = \hbar/e^2 (3\pi^2)^{2/3} n_e^{-1/3} \rho_{xx}^{-1}$ was also calculated for each device under the free-electron approximation~\cite{Ioffe1960}. Here, $k_F$, $l$, and $n_e$ are the \textit{Fermi} wave-vector, the averaged elastic mean free path across the sample dimension, and the charge carrier density extracted from the \textit{Hall} measurement at 10~K, respectively. We note that the metal-to-insulator transition of this material takes places around $\rho_{xx}\sim6\times10^{-3}$ $\Omega$-cm ($k_Fl\sim0.3$), and thus the highest $\rho_{xx}$ used in the study was chosen to be much smaller than this critical value to avoid the transition regime.

\begin{figure*} [t]
      \centering
      \includegraphics[width=16cm]{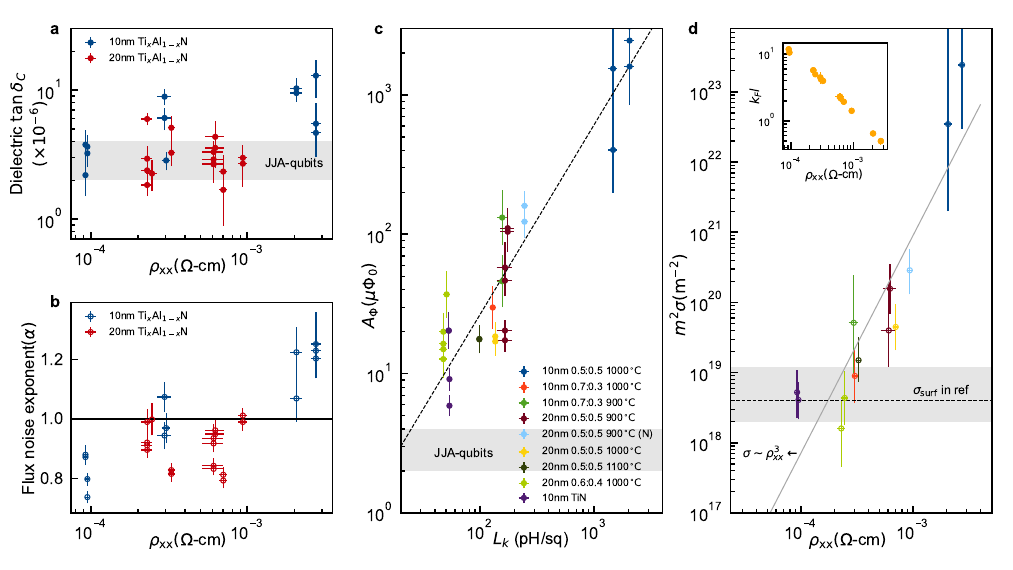}
      \caption{\textbf{The correlation between material disorder and qubit coherence} Values and errors are extracted from experiment coherence times as described in the main text. (a) The dielectric loss of the fluxonium qubits as a function of the $\rho_{xx}$. Shaded area defines the dielectric loss tangent typically measured for fluxonium devices made with JJA. . (b) The extracted flux noise exponent $\alpha$ plotted for all measured devices as a function of the material disorder. (c) The $1/f^\alpha$ flux noise amplitude $\ATtwo$ plotted against the $L_k$ of the inductor wire. Again, the film thicknesses, chemical compositions, and annealing conditions are labeled to each data point accordingly. The dashed line is a guide for the eyes, and the reference values of $\ATtwo$ in fluxonium devices made with JJA are illustrated by the shaded area. The values and errors are obtained using the least squares method under the assumption that noise amplitudes follow a log-normal distribution. (d) The areal density of the phenomenological spin defect ($\sigma$) plotted against the material disorder quantified by $\rho_{xx}$. The inset represents the correlation between measured $\rho_{xx}$ and $k_Fl$. Data points with the same color are averaged from devices on a different wafer but with the same fabrication conditions as labeled in (c), and error bars represent the minimum and the maximum within each dataset. The solid line is the fitting that yields $\sigma \propto \rho_{xx}^{\beta}$ where $\beta = 3.10 \pm 0.28$. The shaded area covers the widely observed areal defect density ($\sigma_{\text{surf}}$) in the referenced superconducting devices under a same frequency integration range from $10^{-4}$ Hz to $10^{9}$ Hz. The dotted line marks $\sigma_{\text{surf}} \sim 4\times 10^{18}$ m\textsuperscript{-2} as a guide for the eyes.} 
      \label{fig:TANNQFig4} 
\end{figure*}

\subsection*{Correlation between disorder and qubit coherence}

We now discuss the qubit coherence properties. The dielectric loss of the qubit set is first plotted against $\rho_{xx}$ (\autoref{fig:TANNQFig4}a). Since the dielectric loss extraction is impacted by the detrimental strongly-coupled TLSs, the error bars are large and have made it challenging to draw a quantitative conclusion. While on a qualitative level, it is indicated that at the weak disorder limit, the dielectric loss $\tan\delta$ of the qubits is not sensitive to the variation of material properties nor the change of capacitance between the inductor wires and the ground planes (\textit{i.e.,} the geometric variation of the wires). Around the qubit operation frequencies, the dielectric loss tangents are found to be comparable to that of fluxonium qubits made with Josephson-junction arrays (JJA) ~\cite{Nguyen2019, Bao2022,Somoroff2023}. At the strong disorder limit, a larger dielectric loss of the qubits is found. The possible source of such increase could be traced to the reduced percolation in thinner films where the segregated grains have sizes comparable to the film thickness, leading to emergent tunneling sites as the detrimental two-level-system defects \cite{Sacepe2008, Carbillet2020, Muller2019}. Plus, as the disordered materials are approaching the superconductor-to-insulator transition, additional dissipative mechanisms have been suggested to appear \cite{Feigelman2018, DeGraaf2020}. Given that a detailed electromagnetic analysis is needed to acquire the exact loss tangent of the \TixAlxN, our findings, nevertheless, have demonstrated the feasibility of employing this material for fluxonium qubits. 

For the $1/f^\alpha$ flux noise, the flux noise exponent $\alpha$ are first plotted for all devices as a function of the material disorder (\autoref{fig:TANNQFig4}b). It is revealed that the flux noise exponents across the two probing frequency regimes (namely, frequencies around the qubit frequencies where $T_1$ traces are measured and frequencies around the spin-echo filter band) take values from $\sim$0.8 to $\sim$1.2. This finding is consistent with a number of reported qubit devices~\cite{Bialczak2007,Bylander2011,Yan2016,Quintana2017}. It is worth noting that $\alpha$ is frequency dependent and can deviate from unity under distinct frequency regimes (see Supplementary Fig.~7 for instance) \cite{Anton2013,Quintana2017,Aquino2022}. Nevertheless, since it is practically challenging to measure the complete noise spectra for all the qubit devices, we have approximated the noise spectra to take a constant $\alpha$ value within the frequency regimes of interest. We argue that such approximation is sufficient for the purpose of this study as the flux noise amplitudes were found to have spanned by at least one order of magnitude, and the impact of fluctuation of $\alpha$ within the frequency regimes can be considered as a secondary effect. Using $\ATtwo$ as an explicit measure of the qubit dephasing properties, it is found that the overall $1/f^\alpha$ flux noise amplitudes are much larger than the JJA-based devices~\cite{Nguyen2019, Bao2022}. In addition, the noise amplitude increases with the increase of the kinetic inductance (\autoref{fig:TANNQFig4}c).

To gain more insight into the physical origins of such a correlation, noise amplitudes were converted to intensive material properties under a phenomenological spin-defect model~\cite{Koch2007, Faoro2008, Bialczak2007, Sendelbach2008, Anton2013, Paladino2014}. Since all of the devices are in the $\lambda > w \gg t$ regime, where the magnetic field distribution is essentially homogeneous \cite{Clem2013}, the total flux variance can be written as $ \langle{\Phi^2}\rangle =\mu_0^2 m_B^2 \sigma/12 \cdot (p/w)$ (see Supplementary Information for more experimental validation)~\cite{Bialczak2007, Anton2013, Braumuller2020}. Here, $\sigma$ is the areal density of the phenomenological spin defects and $m_B = m\mu_B$ is the effective magnetic moment defined by a constant $m$ (depending on the nature of the defect) times the Bohr magneton $\mu_B$. Thus, using $\langle{\Phi^2}\rangle = 2\int_{\omega_1}^{\omega_2}d\omega S_{\Phi}(\omega)$ within the framework of our $1/f^\alpha$ noise approximation\cite{Bylander2011}, the areal density of the phenomenological spin defects $m^2 \sigma=24/\mu_0^2\mu_B^2 \cdot (w/p)\cdot A_{\Phi}^2\omegaref^{\alpha-1}\int_{\omega_1}^{\omega_2}d\omega \cdot \omega^{-\alpha}$

is calculated for each device and plotted versus the material disorder (\autoref{fig:TANNQFig4}d) following the convention  $\omegaref/2\pi=1$ Hz, $\omega_1/2\pi=10^{-4}$ Hz and $\omega_2/2\pi=10^9$ Hz\cite{Koch2007, Anton2013}. Again, owing to the largely homogeneous magnetic-field distribution, we treat the source of the spin defects solely from the bulk of the \TixAlxN inductor wires. 

At the low-disorder limit and assuming $m_B=\mu_B$, the areal spin defect density $\sim 1 \times 10^{18}$ m\textsuperscript{-2} is on par with the universal values measured on superconducting resonators, flux qubits, and superconducting quantum interference devices (SQUIDs) when using the same frequency integration range~\cite{Bialczak2007, Sendelbach2008, Anton2013, DeGraaf2017, Braumuller2020}. With an increase in material disorder, the defect density increases rapidly, indicative of a possible crossover from a surface-limited regime to a disorder-limited regime. Although there are distinctions across different material systems, this result could serve as an additional data set to understand the long-lasting flux-noise problems. More interestingly, by fitting the measured devices, a tentative correlation is acquired and yields $\sigma \propto \rho_{xx}^{\beta}$ with $\beta = 3.10 \pm 0.28\approx 3$. Under the free-electron approximation where $\rho_{xx}^{-1} \propto k_F\cdot(k_Fl)$ and considering the fact that $k_F$ is essentially a constant in our case (Supplementary Table 1), the above relation is thus approximately equivalent to $\sigma \propto (k_Fl)^{-3}$ (see \autoref{fig:TANNQFig4}d inset for the range of $k_F l$).

\section*{Discussion}

We conclude by discussing the implication of the results. First, the physical simplicity and the robustness in achieving a stable dielectric losses are advantageous of disordered \TixAlxN. The compromises, although yet clear for other similar systems, are the large flux noise or the inductive losses intrinsically associated with the disorder. As such, in applications that are flux insensitive \textit{e.g.,} the quasicharge~\cite{Pechenezhskiy2020} and $0$-$\pi$~\cite{Brooks2013} qubits, the introduction of material disorder is a practical approach to engineer for high-impedance circuits. While in the realm of flux-tunable devices such as fluxonium qubits, a strategy from the high-coherence consideration is to further reduce the disorder and deliberately balance between the achievable qubit parameters and the flux-noise sensitivity. 

Secondly, the phenomenological spin defect density is seemingly only dependent on the disorder, irrespective of other material properties. To be specific, as $k_F$ is largely invariant, the intriguing $\sigma \propto (k_Fl)^{-3}$ relation has suggested a possible correspondence between the phenomenological spin defects and the volumetric density of the elastic scattering centers. It is thus reasonable to postulate that either the formation of the spin defects is correlated with that of the scattering defects, or the phenomenological spin defect is a collective manifestation of the scattering processes. It is noted that if the spin defects were assumed to be unpaired electrons with $m_B=\mu_B$ in the strong disorder limit ($\rho_{xx} > 10^{-3}$ $\Omega$-cm), the volumetric spin defect density is comparable to the charge carrier density of the material ($\sim 10^{28}$ m\textsuperscript{-3}). In this case, one might expect the superconductivity to vanish, however, these samples still exhibit robust superconductivity. A tentative rationalization of such counter-intuition is that the superconductivity survives owing to the percolating channels or the strongly-coupled superconducting granules. This bears a resemblance to the granular aluminum~\cite{Deutscher1973, Dynes1981} in which superconductivity is still observed even when the material is prepared with resistivity larger than the critical value. In turn, the flux noise seen by the superconducting granules could be from the high density of the dangling spins residing in the non-superconducting regimes. Such experimental evidence on the correlation between disorder and magnetic spin density could offer additional insights into the mechanisms underlying the phase transition of superconductors~\cite{Sacepe2020} and novel approaches to alleviate such high density of defects.

At a microscopic level, it is thus far unclear whether a specific type of randomly distributed defects (\textit{e.g.,} anion vacancies, anti-sites, interstitial sites, etc.) is to be attributed as the source of such noisy spins. For instance, the fact that the sample set annealed in nitrogen sharing a similar correlation between $\sigma$ and $\rho_{xx}$ with those annealed in argon has indicated the limited impact of the nitrogen vacancies. A background of $\sim$10 at.\% oxygen defects was observed across all samples (see Supplementary Fig.~9) but is difficult to be accounted for the orders of magnitude change in the spin defect density. On the other hand, if the spin defects are actually segregated in the non-superconducting regimes, dangling electron spins from the aliovalent dopants (\textit{i.e.,} Ti-doped aluminum nitride) could be considered as a potential noisy source. Plus, given that the wafers were only annealed for a finite time, the contribution from factors such as non-equilibrium defects, progressive microstructure, and the density of the spinodal segregation boundaries~\cite{Choi2009} are yet to be ruled out.

A potentially helpful and prominent experiment would be the universality check of such relationship for materials with distinct compositions and preparation conditions across a larger parametric space. If a material dependency of the above correspondence fails to exist and the correlation between $\sigma$ and $\rho_{xx}$ (or $k_F l$) still holds, such relation could be more of a fundamental mechanism that governs the disordered electronic systems; while if it does, the two parameters are in principle independent properties, which in turn, can be engineered for a system with simultaneously large kinetic inductance and low flux noise.

\section*{Methods}

\textbf{Device fabrication.} The device fabrication follows and extends from the established approach \cite{Gao2022a}, while a more comprehensive illustration and detailed characterization on the devices are provided in the Supplementary Information. The device fabrication starts from the single-side-polished sapphire wafers sputtered with \TixAlxN films with varied thicknesses and chemical compositions at 300$^{\circ}$C. The chemical compositions were nominal compositions determined from the the sputtering rates of single-layer TiN and AlN films. The as-grown films were coated with a 100 nm-thick SiN\textsubscript{x} hardmask using a plasma-enhanced chemical vapor deposition system (PECVD) at 200$^{\circ}$C. The hardamsk is then lithographically patterned and etched with an inductively-coupled plasma etching system (ICP) into a rectangular-shaped patch, followed by the SC-1 wet etch to remove the exposed \TixAlxN films. After etching, the wafers were sent to the annealing tube of an low-pressure chemical vapor deposition system (LPCVD) to perform thermal treatment. The purpose of the annealing is to introduce phase segregation in the \TixAlxN films, while at the same time the exposed sapphire surfaces are treated. Samples were vertically loaded in the center of the hot zone and the tube was then purged with argon (5N-purity) or nitrogen (5N-purity). A constant flow of argon/nitrogen at 3000 sccm was maintained throughout the anneal at atmospheric pressure. The annealing time was fixed at 30 mins while the annealing temperatures were varied to achieve different material properties. After the thermal processing, the wafers were sent to the sputter chamber again to deposit a layer of 100 nm-thick TiN films. The TiN films were then patterned using the similar hardmask and wet-etch steps to form the backbone of the low-loss quantum circuits. When the patterning was done for both the \TixAlxN patch and the TiN films, the wafer was rinsed in diluted hydrofluoric acid (DHF) to remove all the SiN\textsubscript{x} hardmask layers. A resistivity check was performed at this stage on the \TixAlxN layers to determine the final geometry of the inductor wires. To form an inductor with desired total inductance, the \TixAlxN patch is lithographically patterned with the PMMA resists using a high-resolution e-beam lithography system, and dry etched with the ICP system using a chlorine-based etching recipe. After the patterning, the wafer stack was thoroughly cleaned using organic solvents and the morphology was checked. With the formation of the majority of the fluxonium circuits, the final step is the fabrication of the Manhattan-style Josephson junctions by the shadow-evaporation technique in a high-vacuum e-beam evaporation system. A gentle ion-mill step was added before the evaporation to remove organic residuals and ensure a good galvanic connection between the inductor wires, capacitor pads, and the Josephson junctions. The wafer stacks were then cleaned and diced for the following cryogenic tests. 

\textbf{Cryogenic measurement setup.} Standard heterodyne setups were used to characterize the qubits (the schematics of the measurement setup was given in the Supplementary Fig.~3). Arbitrary waveform generators were used to generate the qubit driving and readout signals, respectively, and then modulated with carrier waves generated by microwave sources. At the signal input, cryogenic attenuators at different stages, low-pass filters, infrared filters, and DC blocks were applied to achieve thermal anchoring and noise suppression. A total of 60 dB and 80dB attenuation were used for the charge driving (XY) and readout inputs, respectively. At the output of the readout signals, an infrared filter, a low-pass filter and three-stage isolators were applied. The signals were then amplified by the high-mobility-electron-transistor amplifiers at the 4K stage and additional amplifiers at the room temperatures. In terms of the flux tuning of the qubits, a DC source and an arbitrary waveform generator were used on two dedicated input lines and combined via a home-made bias-tee at the mixing chamber stage. The DC-flux lines and the fast-flux lines (Z-lines) were equipped with dedicated sets of the RC filters (10 kHz), cryogenic attenuators, home-made copper-powder filters, infrared filters, or low-pass filters to achieve thermal anchoring and noise suppression. A total of 30 dB attenuation were used for the fast-flux lines. Copper, aluminum and $\mu$-metal shields were used surrounding the samples to isolate the samples from environmental noises. 

\textbf{Transport studies.} The transport studies were performed on patterned \textit{Hall}-bar dies patterned along with the qubit wafers. The width of the \textit{Hall}-bar channel is 100 $\mu$m and the length of the channel is 190 $\mu$m. The electrical connections to the sample puck were made by aluminum wedge bonding (see inset of Supplementary Fig.~10 for wiring schematics). The samples were then loaded in a physical-properties measurement system equipped with a tilting sample stage for all the subsequent characterizations. A constant current of 1~$\mu$A was supplied for both longitudinal and \textit{Hall} resistance measurement. 

\textbf{Material characterization} The X-ray diffractometry (XRD) studies including linescans and rocking curves were performed on a high-resolution D8 ADVANCE X-ray diffraction system with optics set up for epitaxial-film studies. The X-ray photo-electron spectroscopy (XPS) studies were performed on a ESCALAB Xi+ XPS system at an incident angle of 60$^{\circ}$. The system was first calibrated with standard samples, and the selective milling between titanium, aluminum, oxygen, and nitrogen was confirmed to be minimal. For the XPS depth-profile studies, the milling area was set to be 2 mm by 2 mm while the analyzing area was concentric with the milling area and set to be 0.4 mm by 0.4 mm. High milling current of the Ar$^+$ was applied, and the beam energy was set to 2000 eV. The spectra were taken after every milling step in a 10 sec interval for the surface region and 60 sec interval for the bulk part. For the transmission-electron microscopy (TEM) studies, the samples were prepared using \textit{in situ} focused-ion-beam liftout and coated with a layer of gold for surface protection. The bright-field images were taken in a aberration-corrected Themis-Z system at 200 keV, and the energy-dispersive spectroscopy (EDS) mappings were taken under scanning-TEM mode with a Super X FEI system. The scanning-electron microscopy (SEM) images were taken in the secondary-electron mode with a GeminiSEM system and the atomic force microscope (AFM) scans were performed on a Jupiter XR AFM system under the conventional AC-tapping mode.   

\section*{Data Availability}
Relevant data supporting the key findings of this study are available within the article and the Supplementary Information file. All raw data generated during the current study are available from the corresponding authors upon request.

\vspace{1cm}
\bibliography{TAN-NQ-Ref-NC}

\section*{Acknowledgements}
We thank the former DAMO Quantum Laboratory team for their technical support during the experimental work at Alibaba Group. We thank Dr. Lev B. Ioffe for the insightful discussion and his comments on the nature of the phenomenological spin defects. We thank Dr. Benjamin Sac{\'{e}}p{\'{e}} for his suggestions on the manuscript and the critical comments on the rigorousness in quantifying the material disorder. We thank Dr. Ioan M. Pop for the insightful discussions on disordered materials and qubit studies. We thank the Westlake Center for Micro/Nano Fabrication for fabrication supports. We thank the Instrumentation and Service Center for Physical Sciences, and the Instrumentation and Service Center for Molecular and Physical Sciences at Westlake University for the material characterization services. R.G., X-Z.M., Fei W., C.D. acknowledge the support from Guangdong Provincial Quantum Science Strategic Initiative (Grant No. GDZX2407001). 

\section*{Author contributions statement}
R.G. and C.D. conceived the central concepts and organized the experiments. R.G. synthesized and characterized the materials and fabricated the devices. H.S. and X-Z.M. characterized the devices and executed the noise spectroscopic measurements. Feng W. analyzed the raw data and designed the automatic data processing protocol. R.G., Feng W., C.D., H.S., X.W., and H-H.Z. performd the analysis and interpretation of the data. J.C., Feng W., T.X., and H-H.Z. designed the qubit parameters and circuit layout. H.D. measured the resonator devices. Z.S. assisted the cryogenic measurement setup. C.Z., X-H.M., M.Y., Fei W. assisted the material characterization and device fabrication. R.G., Feng W., H.S., C.D., and Y.S. wrote the core of the manuscript. All authors contributed to the understanding of the results and the writing of the manuscript.

\section*{Competing interests statement}

The authors declare no competing interests.

\end{document}


\vspace {1cm}

\title{\LARGE Supplementary Information \\
\Large The effects of disorder in superconducting materials on qubit coherence}

\author{Ran Gao}
\thanks{These authors contributed equally to this work.}
\affiliation{Quantum Science Center of Guangdong-Hong Kong-Macao Greater Bay Area, Shenzhen 518045, China}
\affiliation{Z-Axis Quantum, Hangzhou, China}

\author{$^{\!\!,\;\textcolor{blue}\dagger}$ Feng Wu}
\thanks{These authors contributed equally to this work.}
\affiliation{Zhongguancun Laboratory, Beijing, China}

\author{Hantao Sun}
\thanks{These authors contributed equally to this work.}
\affiliation{China Telecom Quantum Information Technology Group Co., Ltd., Hefei 230031, China}

\author{Jianjun Chen}
\affiliation{Xinxiao Electronics Inc., Hangzhou 310018, China}

\author{Hao Deng}
\affiliation{International Center for Quantum Materials, Peking University, Beijing 100871, China}

\author{Xizheng Ma}
\affiliation{Quantum Science Center of Guangdong-Hong Kong-Macao Greater Bay Area, Shenzhen 518045, China}

\author{Xiaohe Miao}
\affiliation{Instrumentation and Service Center for Molecular and Physical Sciences and Research Center for Industries of the Future, Westlake University, Hangzhou 310030, China}

\author{Zhijun Song}
\affiliation{Shanghai E-Matterwave Sci \& Tech Co., Ltd., Shanghai 201100, China}

\author{Xin Wan}
\affiliation{Zhejiang Institute of Modern Physics and Zhejiang Key Laboratory of Micro-nano Quantum Chips and Quantum Control, Zhejiang University, Hangzhou 310027, China}

\author{Fei Wang}
\affiliation{Quantum Science Center of Guangdong-Hong Kong-Macao Greater Bay Area, Shenzhen 518045, China}
\affiliation{Z-Axis Quantum, Hangzhou, China}

\author{Tian Xia}
\affiliation{Huaxin Jushu Microelectronics Co., Ltd., Hangzhou 310052, China}

\author{Make Ying}
\affiliation{EXTEC Inc., Hangzhou 310024, China} 

\author{Chao Zhang}
\affiliation{Instrumentation and Service Center for Physical Sciences, Westlake University, Hangzhou 310024, Zhejiang, China}

\author{Yaoyun Shi}
\affiliation{Z-Axis Quantum, Hangzhou, China}

\author{Hui-Hai Zhao}
\affiliation{Zhongguancun Laboratory, Beijing, China}

\author{Chunqing Deng}
\email{gaoran@quantumsc.cn; \\ dengchunqing@quantumsc.cn}
\affiliation{Quantum Science Center of Guangdong-Hong Kong-Macao Greater Bay Area, Shenzhen 518045, China}
\affiliation{Z-Axis Quantum, Hangzhou, China}

\maketitle

\section{Device fabrication and qubit morphology}
\begin{figure*}[t]
      \centering
      \includegraphics[width=8.6cm]{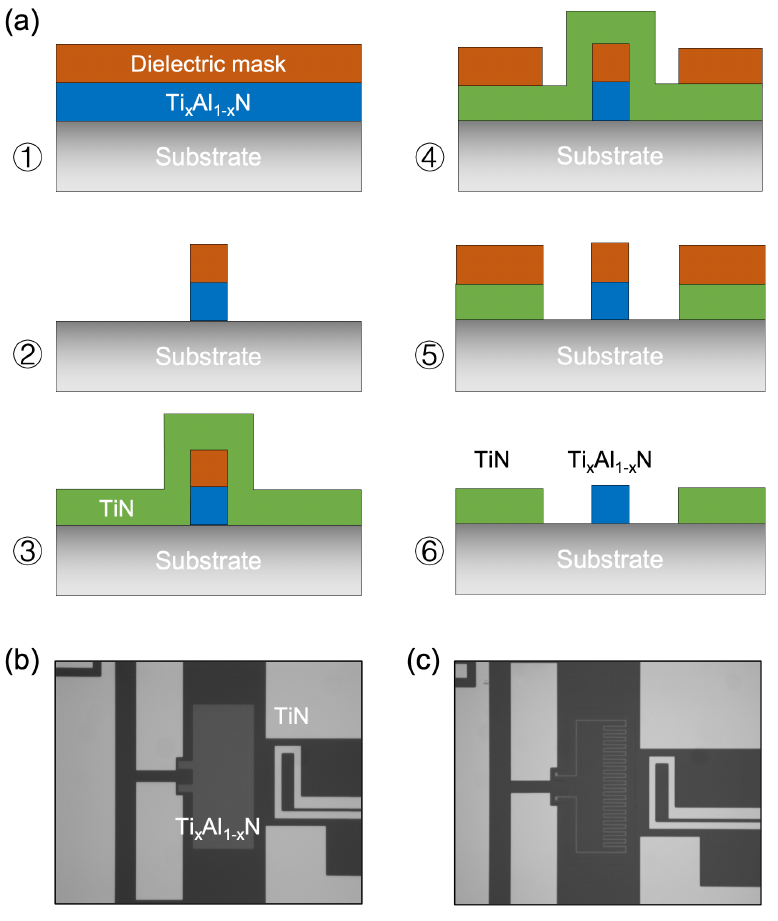}
      \caption{(a) The fabrication process flow of the Ti\textsubscript{x}Al\textsubscript{1-x}N-based fluxonium qubits. Step 1: Dielectric hardmask coating on the Ti\textsubscript{x}Al\textsubscript{1-x}N films. Step 2: Hardmask patterning followed by wet etching of Ti\textsubscript{x}Al\textsubscript{1-x}N using the SC-1 solution and annealing. Step 3: Deposition of the TiN films. Step 4: Hardmask deposition and patterning. Step 5:  Wet etching the TiN films. Step 6: Hardmask stripping by diluted hydrofluoric solutions. (b) The optical image of the Ti\textsubscript{x}Al\textsubscript{1-x}N patch and the rest of the circuit after the hardmask striping (\textit{i.e.,} Step 6). (c) After Step 6, the wafers were re-coated with the e-beam resists and patterned to form the inductor wires by dry etch. The optical image of the patterned Ti\textsubscript{x}Al\textsubscript{1-x}N inductor wires and the other circuits before the junction fabrication.}
      \label{fig:TANNQFigS1} 
\end{figure*}

The fabrication process flow is provided (\autoref{fig:TANNQFigS1}) and details are explained in Methods. This flow is designed to minimize the process induced artifacts, and we perform characterization at each key step as a precautious measure to ensure the qubit yield and the device quality. It is worth noting that, the wet etching process at step 5 inevitably attack the underneath Ti\textsubscript{x}Al\textsubscript{1-x}N patch from the sidewalls and create rough edges. However, since the dry etch is applied in the following steps with a larger etching area than the footprint of the Ti\textsubscript{x}Al\textsubscript{1-x}N patch (see the optical image in the main text), rough edges formed in the wet-etch step can be fully removed by the gas mixture.

\begin{figure*}[b]
      \centering
      \includegraphics[width=8.6cm]{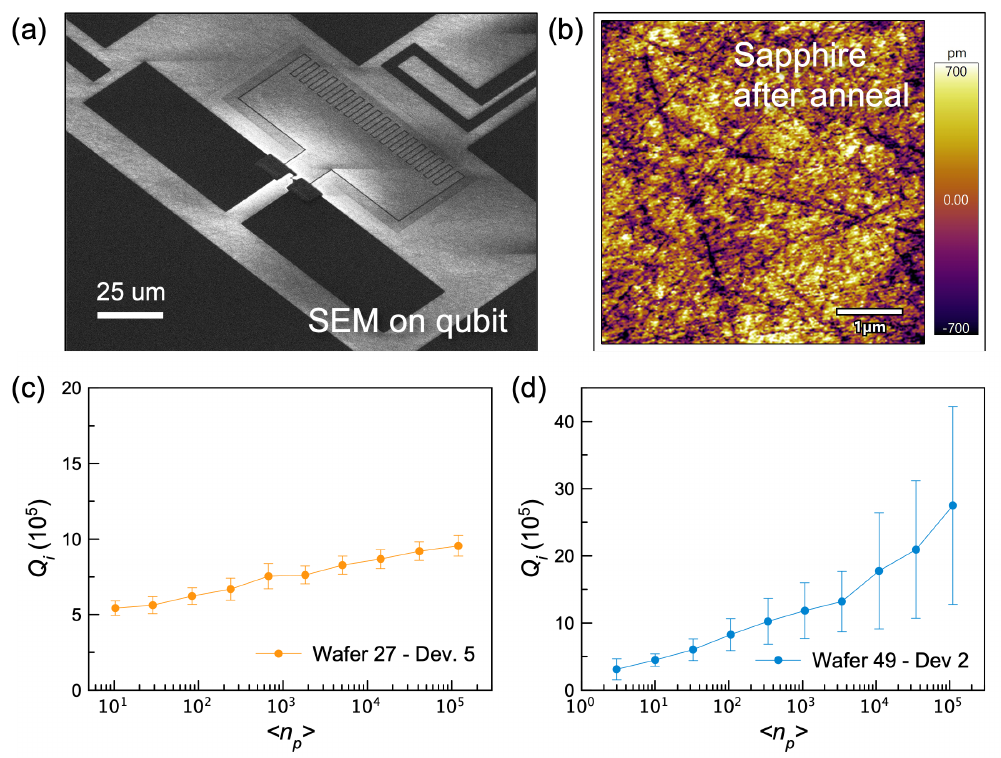}
      \caption{(a) The scanning electron microscopy (SEM) image of the fluxonium qubits. Some artifacts on the image are due to the electron charging on the sapphire substrates. The different contract at the regime surrounding the inductor wire is caused by a different sapphire surface properties after the dry-etch process. (b) The sapphire surface qualities were examined after the annealing of the Ti\textsubscript{x}Al\textsubscript{1-x}N patches and before the deposition of titanium nitride. (c-d) The quality factors of resonator devices fabricated using 10-nm thick and 20-nm thick Ti\textsubscript{0.5}Al\textsubscript{0.5}N films with TiN ground planes. }
      \label{fig:TANNQFigS2} 
\end{figure*}

In terms of the eventual qubit devices, the scanning electron microscopy (SEM) of the qubits revealed a clean morphology after the formation of the Josephson junctions (\autoref{fig:TANNQFigS2}a). Together with the optical images and the atomic force microscopy (AFM) scans, no observable fabrication artifacts were introduced by the multi-step processes. We do find some residual organics around the Josephson-junction area that are commonly seen in superconducting qubits fabricated using conventional lift-off approaches. These artifacts could be further optimized in future works. 

Since the process flow incorporates a high-temperature annealing step that potentially brings in uncertainties in the qubit qualities, AFM scans on the sapphire surfaces were performed and we found that the surface roughness is essentially unchanged (\autoref{fig:TANNQFigS2}b). No surface reconstruction and atomic steps were found after annealing since the sapphire surfaces have experienced one round of Ti\textsubscript{x}Al\textsubscript{1-x}N deposition and wet etch. The sharp and long trenches could be the surface manifestation of the internal crystal structures or defects after the annealing. Nevertheless, a sanity check was performed on TiN resonators fabricated on the annealed sapphire substrates, and the internal quality factors are the same as those fabricated on the as-received sapphires reported before~\cite{Gao2022}. 

Further examination on the quality of the Ti\textsubscript{x}Al\textsubscript{1-x}N materials and the fabrication processes were also performed on microwave resonators. The resonator designs and testing protocols follow the established approaches~\cite{Gao2022a}. Here, two exemplary samples are presented (\autoref{fig:TANNQFigS2}c-d). At the low-photon-number regime, the internal quality factors $Q_i$ of the resonators range from $\sim$300k to $\sim$500k, and showed photon-number dependency. At the low-photon-number regime, the quality factors typically take a value between 10 million and 20 million. This behavior is reasonably consistent amount samples with kinetic inductance smaller than 1~nH/sq.

\section{Cryogenic measurement setup}
      
\begin{figure*} [b]
      \centering
      \includegraphics[width=9.6cm]{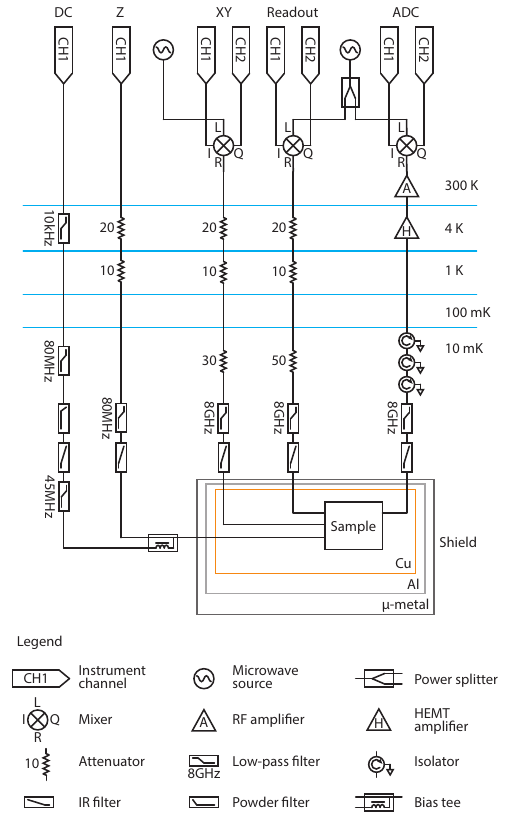}
      \caption{The schematics of the cryogenic measurement setup. The attenuators are labeled with the corresponding attenuation level in the unit of decibel (dB). The commercial low pass filters are labeled with their cutoff frequencies.}
      \label{fig:TANNQFigS3} 
\end{figure*}

A schematic of the cryogenic measurement setup is provided (\autoref{fig:TANNQFigS3}), and the details of the setup are discussed in the Methods section. 

\clearpage
\section{Decoherence analysis}

\begin{figure} [b]
      \centering
      \includegraphics[width=8.6cm]{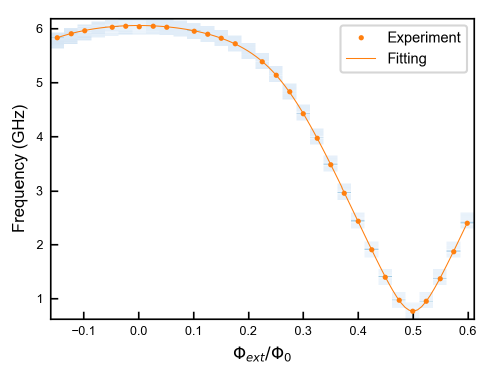}
      \caption{A typical fluxonium spectrum versus external flux. The qubit transition frequencies (orange markers) are extracted by peak finding in the qubit population signal versus the qubit excitation frequency. For this particular qubit, a fit to the fluxonium Hamiltonian (orange line) yields $E_C/h=1.39~\GHz$, $E_{J}/h=4.10~\GHz$, and $E_L/h=0.85~\GHz$.}
      \label{fig:TANNQFigS4} 
\end{figure}

The Hamiltonian of a fluxonium qubit can be written as
\begin{equation}
    \hat{H}= 4 E_C \hat{n}^{2}+\frac{E_L}{2}\left(\hat{\varphi} + \frac{\Phi_{\text{ext}}}{\varphi_0}\right )^{2}-E_J \cos\hat{\varphi}\ , \label{eq:qubit_hamiltonian}
\end{equation}
where $E_C$, $E_L$ and $E_J$ are the charging, inductive, and Josephson energies, $\Phiext$ is the external flux, $\hat{\varphi}$ is the operator of the phase across the Josephson junction, and $\hat{n}$ is the charge operator in the unit of $2e$. The $\hat{\varphi}$ and $\hat{n}$ are conjugate variables. $\varphi_0 = \Phi_0/2\pi$ is the reduced flux quantum. We measure fluxonium's spectra using a flux-pulse protocol approach~\cite{Sun2023} in a large (over half-flux-quanta period) external flux range. Qubit parameters are extracted from fitting to the fluxonium Hamiltonian. One set of typical fluxonium spectrum data and a fit is shown (\autoref{fig:TANNQFigS4}).

\onecolumngrid

\begin{figure} [t]
      \centering
      \includegraphics[width=16cm]{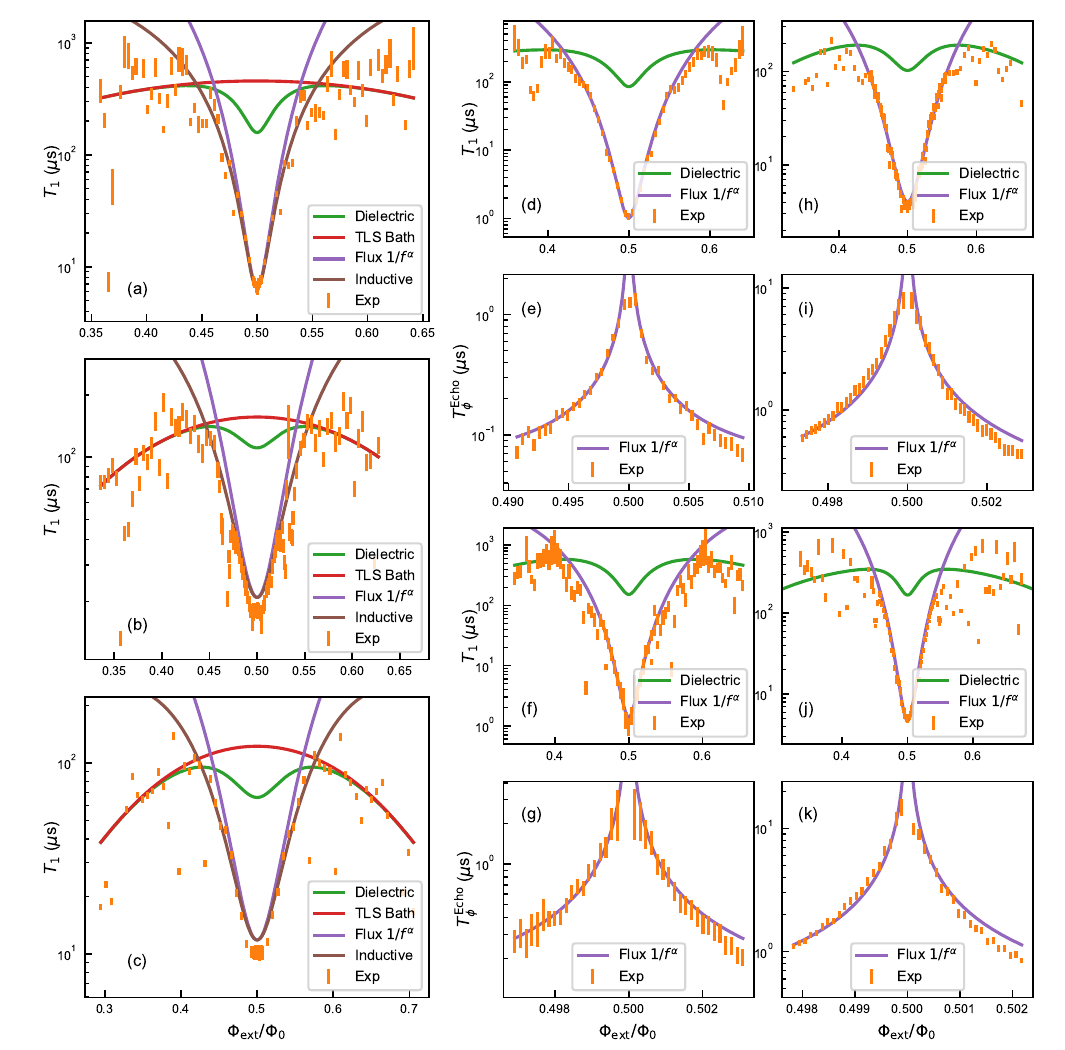}
      \caption{Possible relaxation models for three typical fluxonium qubits with different material disorder including (a) D4\_2, (b) D5\_2, and (c) D7\_2. The markers are the measured $T_1$ and $T_{\phi}^{\text{Echo}}$ values and the lines represent different models. Note that for representation clarity, we did not add total simulated curves for this particular figure. We have also included additional exemplary devices ((d)(e) D1\_3, (f)(g) D8\_2, (h)(i) D9\_1, (j)(k) D11\_2) as a complementary to Fig.~2 in the main text to better illustrate the model applicability. }
      \label{fig:TANNQFigS_CompareFit} 
\end{figure}
\twocolumngrid

In this study, we use a decoherence model assuming two dominant decoherence channels, the $1/f^\alpha$ flux noise and dielectric loss, to describe both the $T_1$ and $T_{\phi}$ data. To obtain a reasonable estimation of the values of the noise parameters, we fit our decoherence model to the data with the following equations:
\begin{align}
 T_1 &= 1 / (\Gamma_1^{\textrm{diel}} + \Gamma_1^{\textrm{flux}}), \label{eq:fit_t1}\\
 T_{\phi}^{\textrm{Echo}} &= 1 / \Gamma_{\phi}^{\textrm{flux}}, \label{eq:fit_tphi}
\end{align}
where the values of $T_1$ and $T_{\phi}^{\textrm{Echo}}$ were extracted from $P(t)$ curves from the measurements. The decay rates, $\Gamma_1^{\textrm{diel}}$, $\Gamma_1^{\textrm{Flux}}$ and $\Gamma_{\phi}^{\textrm{Flux}}$ , follow the definitions in the main text. The parameters $A_{\Phi}, \alpha, \tan\delta_C$ are extracted by the same analysis procedures for every measured device. We acknowledge that although we can extract the error distributions of $T_1$ and $T_{\phi}^{\textrm{Echo}}$ from the $P(t)$ curves, the error from the fluctuation of the system is unknown and difficult to be quantified. For instance, strongly-coupled TLSs are detrimental to the coherence at high frequencies. To account for the system fluctuations and better extract the fitting error systematically, the errors were estimated by assuming a normal distribution of the logarithmic decoherence time $\ln(T)$.

It is worth noting that what the fitting method did is essentially extracting an averaged flux noise exponent $\alpha$ across two distinct frequency regimes, \textit{i.e.}, frequencies around $\Gamma_1^{\textrm{diel}} + \Gamma_1^{\textrm{flux}}$ at several gigahertz, and around $\Gamma_{\phi}^{\textrm{flux}}$ below megahertz. As such, since we care about the impact of noises near qubit operational frequencies, the frequency span has sufficiently covered our target of interest. Together with the fact that the flux noise exponents extracted are very close to unity, and that the flux noise amplitudes in our system have spanned by two orders of magnitudes (Fig. 4c in the main text), we argue that the above decoherence analysis is sufficient for the purpose of this study.

\begin{figure} [t]
      \centering
      \includegraphics[width=8cm]{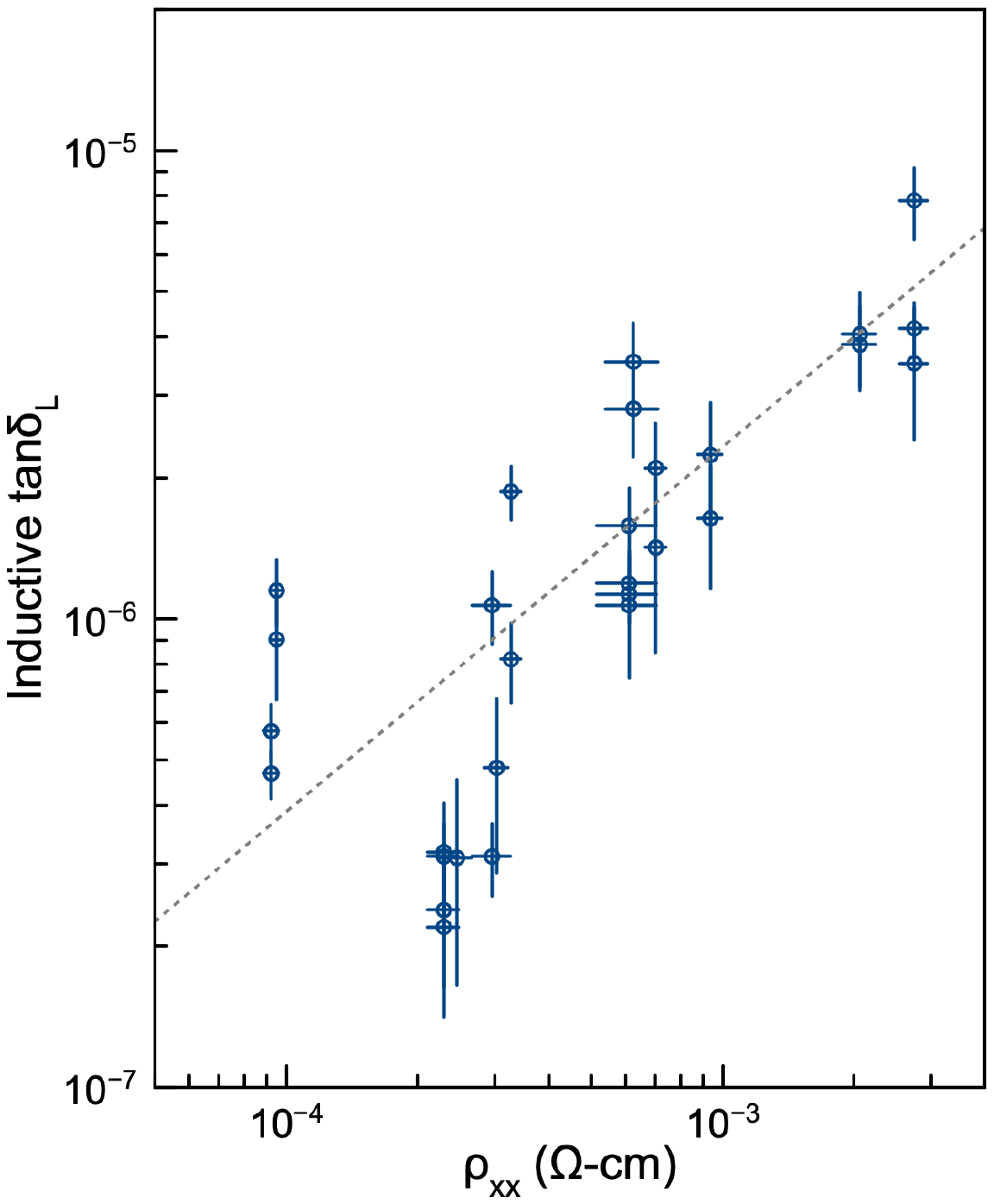}
      \caption{The extracted inductive loss using Supplementary Eq. (\ref{eq:t1_inductive}) as a function of the material disorder. The dotted line serves as a guide for the eyes. The inductive losses are following the same trend as the flux noise amplitudes. As discussed in texts, with the current dataset, we cannot distinguish the $1/f^{\alpha}$ noise and the inductive losses.}
      \label{fig:TANNQFigS7} 
\end{figure}

It is also acknowledged that other type of noises, like dielectric loss from a TLS bath~\cite{Sun2023} and inductive loss, are also possible loss channels in this system. We can write the relaxation rates from these two sources as
\begin{align}
    \Gamma_1^{\text{TLS}} = & \frac{\hbar \omega_{01}^{2}}{4 E_C}  \abs{\mel{0}{\hat{\varphi}}{1}}^2 \tan\delta_{\text{TLS}}, \label{eq:t1_tls}  \\
    \Gamma_1^{\text{ind}} = &  \frac{2E_L}{\hbar}\abs{\mel{0}{\hat{\varphi}}{1}}^2\tan\delta_L\coth\left( \frac{\hbar \omega_{01}}{2k_B\Teff}\right), \label{eq:t1_inductive}
\end{align}
where $\tan\delta_\text{TLS}$ and $\tan\delta_\text{L}$ are the TLS and inductive loss tangent, respectively. A comparison between multiple models is provided (\autoref{fig:TANNQFigS_CompareFit}). The dielectric loss is dominated when the qubit is biased far away from the flux-frustration point, where the qubit frequency is in the gigahertz range, and the temperature coefficient $\coth\left( \hbar \omega_{01}/(2k_B\Teff)\right) \approx 1$. Therefore, we cannot distinguish between the dielectric loss and the TLS model based in this study. That said, we have adopted a more phenomenological term, \textit{i.e.}, dielectric loss, to describe the system other than using “TLS loss” that inherently requires additional assumptions. 

On the other hand, the inductive loss is also a plausible explanation for the $T_1$ spectra around the flux-frustration spots. As stated in the main text, we note that the inductive loss model with frequency dependent loss tangent and the $1/f^{\alpha}$ flux noise model are equivalent. We have extracted the inductive loss tangents and plotted as a function of material disorder (Supplementary~\autoref{fig:TANNQFigS7}). One can find that, the overall trend is very similar to that of the flux noise amplitudes. Nevertheless, we choose the $1/f^{\alpha}$ flux noise model as it naturally lead to dephasing, and thus, we did not need to explain the decoherence properties with additional variables. All said, we have adopted dielectric loss and $1/f^\alpha$ flux noise as the major decoherence sources in our analysis to both minimize the variables in studying such complicated material systems and to make quantitative cross-sample comparison.

\section{The flux noise spectrum}
\begin{figure} [b]
      \centering
      \includegraphics[width=8.6cm]{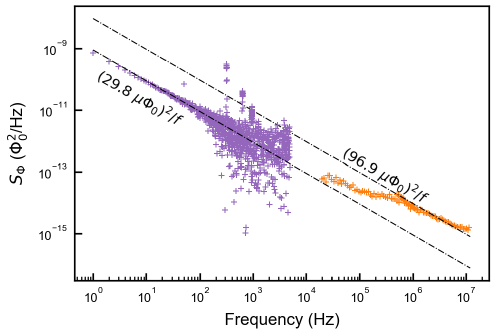}
      \caption{The power spectral density (PSD) of flux noise measured on D7\_3. The low frequency part of the PSD (purple) is obtained from the PSD of the binary time series from single-shot readout with the white sampling noise subtracted. The high frequency part (orange) is obtained from spin-locking measurements. The spin-locking measurements are performed on two $\Phiext$ points with different flux-noise sensitivity $\partial \omega_{01}/\partial \Phiext$ to cover the frequency range of the presented data.}
      \label{fig:TANNQFigS8} 
\end{figure}

As discussed in the main text, the flux noise exponent could have frequency-dependent variations. Although we have extracted $\alpha$ by fitting both $T_1$ and $T_{\phi}$ data, it is also interesting to find out the exact noise spectra and examine if there are fine features of $\alpha$ for such material system. 

The precise form of the flux noise spectrum across a wide range of frequencies can be probed by a combination of the direct sampling and the spin-locking measurements~\cite{Yan2012,Yan2013,Quintana2017}. The power spectral density (PSD) of the flux noise from one specific device D7\_3 is given (\autoref{fig:TANNQFigS8}). Both measurements are performed at $\Phiext\ne \Phi_0/2$ where the qubit frequency is first order sensitive to the flux noise. 

To obtain the flux noise PSD in the quasi-static regime, we use a Ramsey interference experiment~\cite{Yan2012} to sample the frequency fluctuation of the qubit. We repeat the operation sequence, where the qubit undergoes Larmor precession with a fixed time followed by single-shot readout, yielding a binary time series $\{b_n\}$ of the qubit state. The bilateral PSD $S(\omega)= \int_{-\infty}^{\infty}d\tau e^{i\omega \tau}\langle b(\tau)b(0)\rangle = |Z_{k = t_s N\omega/2\pi}|^2/(N/t_s)$ with $\{Z_k\}$ the discrete Fourier transform of $\{b_n\}$, subtracted by a white statistical sampling noise $S_w = \langle b \rangle (1-\langle b \rangle) t_s $, is connected to the flux noise PSD by $S_{\Phi}(\omega)=(S(\omega)-S_w)/( a\tau_0\partial \omega_{01}/\partial\Phiext)^2$, where $2a$ is the readout visibility and $\tau_0$ is the precession time. The frequency range of this method is $[1/2t_s N, 1/2t_s]$, where we use the sampling interval $t_s=10^{-4}\ \text{s}$ and the total number of cycles $N=10^4$. At the qubit operation point, we have $2a = 0.5$, $\tau_0 = 100$~ns, $\partial \omega_{01}/\partial\Phiext = 2\pi\times 2.16~\text{GHz}/\Phi_0$, and $\langle b \rangle = 0.55$. We take the ensemble average of the PSD from $1.5\times 10^4$ time traces to improve the data accuracy.

For the flux noise spectroscopy in the kilohertz to megahertz regime, we perform spin-locking measurements similar to those described in Ref.~\cite{Yan2013}. To obtain the noise PSD at a given frequency, the SL-5a spin-locking pulses are applied with Rabi frequency $\Omega_R$. The rotating frame relaxation rate follows $\Gamma_{1\rho} =\Gamma_{\nu}(\Omega_R)+\Gamma_1/2$, where $\Gamma_1$ is obtained from independent $T_1$ measurements at the same flux position. The decay rate $\Gamma_{\nu}$ is related to the longitudinal noise PSD by $S_Z(\omega)=2\Gamma_{\nu}(\omega)$, and the flux noise by $S_\Phi(\omega)= 2\Gamma_{\nu}(\omega)/(\partial \omega_{01}/\partial \Phiext)^2$. To extend the covered frequency range of this spectroscopy method which requires the noise in the weak coupling regime ($\Omega_R \gg \Gamma_{1\rho}$), we sweep the Rabi frequency at two different flux position corresponding to different sensitivity $\partial \omega_{01}/\partial\Phiext = 2\pi\times 2.16 \text{ and } 0.32~\text{GHz}/\Phi_0$ respectively.

To our best knowledge, this is the first measurement of the flux noise PSD in a device made from disordered superconductors. Although much are to be explored and understood for this system, it can be clearly told from the flux noise PSD that the entire spectra deviates from the exact $1/f$ form. In particular, the PSD at low frequencies between 1-100~Hz and at high frequencies between 1-10~MHz are close to the $1/f$ form but with different amplitudes. At intermediate frequencies around 5-500~kHz, the noise PSD is $\sim 1/f^{0.55}$, bridging the low-frequency and the high-frequency part of the PSD. In the ideal case, one is expected to measure the PSD for all the devices to extract precise noise amplitudes associated with different materials. However, such experiments are practically challenging, and we have chosen to make an approximation that the PSD of our devices take the form of $1/f^{\alpha}$ across the frequency regimes of interest.

\section{Tuning the disorder and material characterization}

\subsection{Structural and chemical characterization}

\begin{figure} [b]
      \centering
      \includegraphics[width=8.6cm]{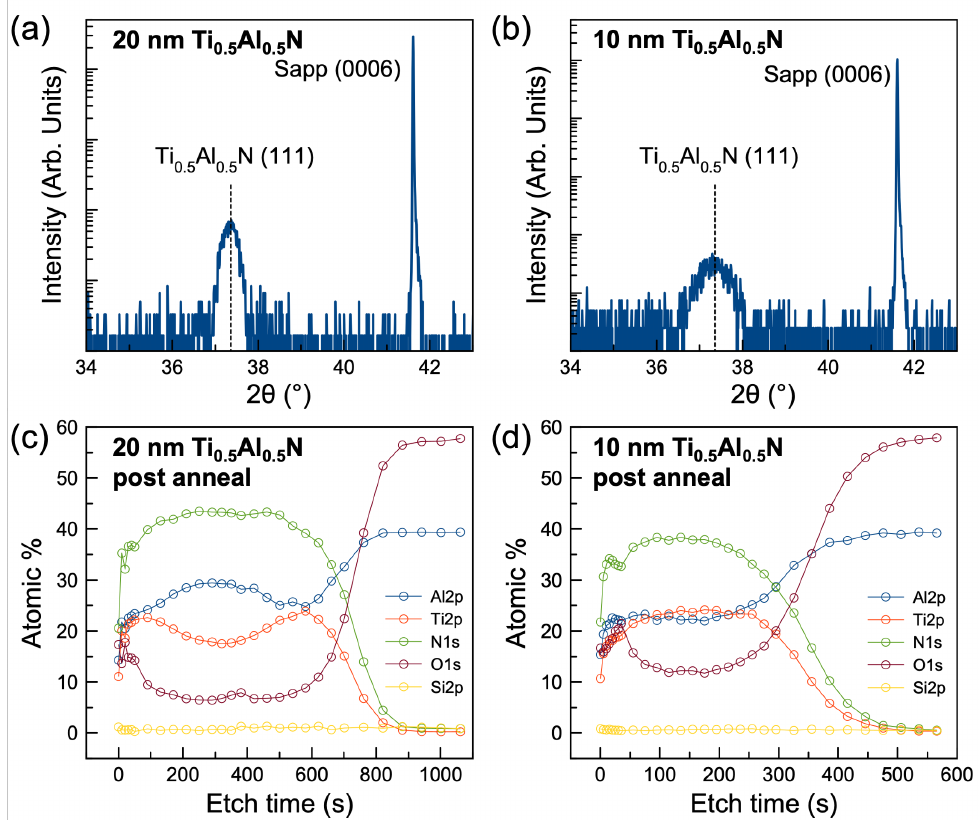}
      \caption{(a-b) The X-ray diffraction line scans of two exemplary Ti\textsubscript{x}Al\textsubscript{1-x}N films grown on the sapphire c-cut substrates. The dashed lines mark the expected peak position of fully-relaxed Ti\textsubscript{0.5}Al\textsubscript{0.5}N films. (c-d) The X-ray photoelectron spectroscopy depth-profile studies on the film stacks after annealing at 1000$^{\circ}$C for 30 mins with a 100 nm-thick SiN\textsubscript{x} hardmask capping layer. }
      \label{fig:TANNQFigS9} 
\end{figure}

Here we perform a set of fundamental material characterization on the set of Ti\textsubscript{x}Al\textsubscript{1-x}N films used for qubit studies. The epitaxial nature of the Ti\textsubscript{x}Al\textsubscript{1-x}N films were revealed by performing standard on-axis theta-2theta X-ray scans, and two exemplary film sets are provided (\autoref{fig:TANNQFigS9}a-b). The (111)-diffraction peaks of the 20 nm-thick and 10 nm-thick films are at the expected positions calculated from the lattice parameters of the TiN and cubic AlN, consistent with the previous results and thicker films~\cite{Gao2022a}. 

\begin{figure*} [t]
      \centering
      \includegraphics[width=12.9cm]{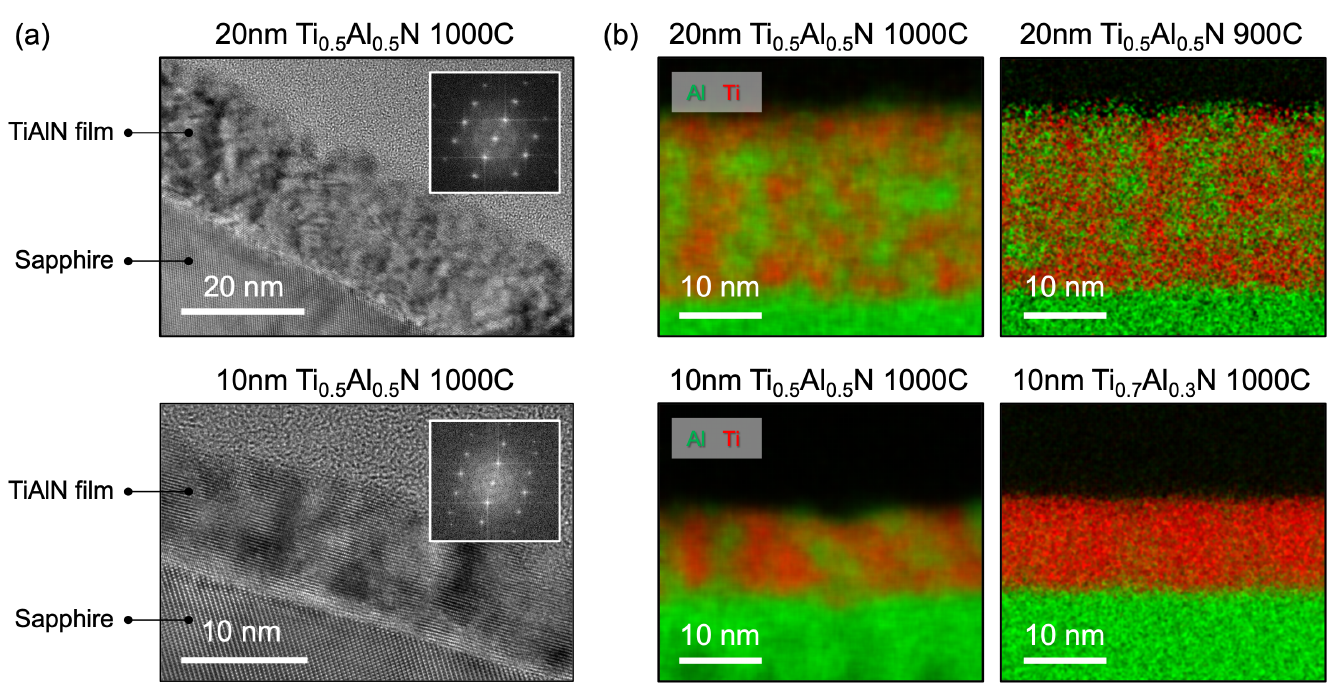}
      \caption{(a) The bright-field TEM scans on two Ti\textsubscript{0.5}Al\textsubscript{0.5}N film stacks with different thicknesses. The inset gives the FFT of the films. (b) The EDS scans of four different film stacks with varied film thicknesses, chemical compositions, and annealing conditions. Here aluminum and titanium cations are mapped with green and red color, respectively. }
      \label{fig:TANNQFigS10} 
\end{figure*}

X-ray photoelectron spectroscopy (XPS) depth-profile studies were performed after the annealing (1000$^{\circ}$C for 30 mins) and stripping of the SiN\textsubscript{x} hardmask, and the atomic percentage of the atoms are plotted against the ion etch time (\autoref{fig:TANNQFigS9}c-d). There are several findings from this study and we elaborate in details here. First, there are roughly $5-10\%$ of oxygen in the 20 nm-thick films, and even more in the 10 nm-thick films. The introduction of oxygen could be from the deposition stage due to the relatively high base pressure of the sputtering chamber ($\sim 8\times10^{-7}$ torr), or during the high temperature annealing. The impact of oxygen on the qubit coherence is hitherto unclear, while we argue that it is not the dominant source of the flux noises. Although the flux noises that are material sensitive, the areal density of the phenomenological spin defects span across four orders of magnitudes (see data in the main text), much greater than the range of changes we observed for the concentration of oxygen defects. Plus, for samples with different chemical compositions, the oxygen concentration is roughly the same but the flux-noise amplitudes have considerably large variation, suggesting that the oxygen defects have yet become the limiting factor of the flux noise in our devices. While for the dielectric losses, it could be true that the oxygen defects are detrimental, as the loss tangent of the 10 nm-thick Ti\textsubscript{0.5}Al\textsubscript{0.5}N films, where the oxygen concentration is as high as $\sim12\%$, is much larger. Using a deposition chamber with better base pressure control for targeting a low oxygen-defect concentration would be worthwhile in the future as a sanity check. 

Besides the segregation at a microscopic level, we have also found certain level of mesoscopic distribution of titanium and aluminum atoms along the out-of-plane direction, particularly for the 20 nm-thick films. It is worth noting that, the exact vertical distribution of atoms depends on whether the film stacks are capped with the dielectric hardmask. In other words, this is suggesting that the vertical chemical profile could be associated the interfacial stress at the film boundaries where a gradient of such stress across the film thickness drives the diffusion of the cation atoms. This finding is consistent with early studies on the impact of strain on the diffusion of atoms~\cite{Knutsson2013, Greczynski2019}. Thus, from a consistency consideration, the process flow is designed such that the Ti\textsubscript{x}Al\textsubscript{1-x}N films were all annealed with a 100 nm-thick SiN\textsubscript{x} hardmask capping layer. Last but not the least, due to the fact that the Ti\textsubscript{x}Al\textsubscript{1-x}N films are annealed in contact with the SiN\textsubscript{x} hardmask, we specifically examined the concentration of silicon in the films since the silicon atoms could potentially act as aliovalent dopants. It can be revealed from the scans that the concentration of silicon is negligible, indicative of a sharp and diffusion-less boundary between the film and the hardmask. 

\subsection{Microscopic structures}

With a better understanding on the overall structures and the chemical compositions of the film stacks, we turn to a detailed and microscopic characterization using the transmission electron microscopy (TEM) and the incorporated energy-dispersive spectroscopy (EDS) studies. Here, the bright-field TEM images on the two exemplary samples are given (\autoref{fig:TANNQFigS10}a). Again, the crystallinity and the epitaxial nature of the samples are confirmed by the fast-Fourier-transform (FFT) insets, and the diffraction contrasts within the samples are attributed to the phase segregation and the local strains~\cite{Knutsson2013, Calamba2021, Calamba2019}. These inhomogeneity can also be reflected in the FFT of the films where a smeared background is overlapped on top of the \textit{Bragg} spots (\autoref{fig:TANNQFigS10}a, insets). 

To investigate the phase segregated structures under a range of preparation conditions, we proceed to EDS scans and map the spacial distribution of the titanium and aluminum cations. Here, four exemplary film stacks are provided to illustrate the microscopic structural evolution under varied film thicknesses, chemical compositions, and annealing conditions, used for the disorder tuning (\autoref{fig:TANNQFigS10}b). First, reducing the film thickness from 20 nm to 10 nm results in a larger segregated grains with sizes comparable to the thickness of the films. In particular for the Ti\textsubscript{0.5}Al\textsubscript{0.5}N films, the reduction of film thickness results in more isolated grains and manifests as a sharp increase in the film resistivity. This evolution can be also understood as a percolation system with six-fold coordination going from 3D to 2D accompanied by a reduction in the percolation threshold from $\sim 30\%$ to $\sim 50\%$. Indeed, if the concentration of titanium in the 10 nm-thick films increases from $\sim50\%$ to $\sim70\%$, the total resistivity of the films greatly reduce and become comparable to those measured in the 20 nm-thick films with $\sim50\%$ of titanium. For different annealing temperatures, in addition, the difference in microstructures is nothing but a consequence of a diffusion-limited process. In other words, the lower annealing temperature reveals finer grains as the diffusion is weak, while an increase in the temperature generates more segregated grains at a larger scale. As being discussed earlier, in thinner films, the increase of annealing temperature give rise to a stronger material disorder after the segregation regimes are grown to have sizes comparable to the film thicknesses. At the opposite end, in thicker films where the scenario remains to be 3D, the higher annealing temperature results in wider conduction channels, better inter-grain connectivity, and higher inter-grain conductivity (due to the increased titanium concentration) that all contribute to a smaller overall material disorder. 

\subsection{Electrical transport characterization}

\begin{table*}
\centering
\caption{The material and device parameters of measured devices. $k_F$ and $l$ are extracted from the transport measurements, while the London penetration depth $\lambda$ is calculated from the kinetic inductance $L_k\approx \mu_0\lambda^2/t$ assuming a homogeneous penetration in the material. The extracted dielectric loss $\tan{\delta}_C$, $1/f^\alpha$ noise amplitude $\ATtwo$ at 1 Hz, flux noise exponent $\alpha$, qubit parameters ($E_J$, $E_C$, $E_L$), and the effective temperature $T_{\text{eff}}$ extracted from thermal population of the qubits are also given. The annealing were performed in argon environment unless specifically noted otherwise. }
\label{tab:table1}
\begin{tabularx}{510pt}{cccccccccccccc>{\centering\arraybackslash}p{25pt}>{\centering\arraybackslash}p{25pt}}
\hline\hline
Device & Ti:Al   & \begin{tabular}[c]{@{}c@{}}Ann. \\ Cond.\end{tabular} & \begin{tabular}[c]{@{}c@{}}$t$\\ (nm)\end{tabular} & \begin{tabular}[c]{@{}c@{}}$k_F$\\ (nm\textsuperscript{-1})\end{tabular} & \begin{tabular}[c]{@{}c@{}}$l$\\ (nm)\end{tabular} & \begin{tabular}[c]{@{}c@{}}$\lambda$\\ ($\mu$m)\end{tabular} & \begin{tabular}[c]{@{}c@{}}$w$\\ ($\mu$m)\end{tabular} & \begin{tabular}[c]{@{}c@{}}$p$\\ ($\mu$m)\end{tabular} & \begin{tabular}[c]{@{}c@{}}$\tan\delta_C$\\ ($\times10^{-6}$)\end{tabular} & \begin{tabular}[c]{@{}c@{}}\ATtwo\\ ($\mu\Phi_0$)\end{tabular} & \begin{tabular}[c]{@{}c@{}}$\alpha$\\ \end{tabular} & \begin{tabular}[c]{@{}c@{}}$E_C/h$\\ (GHz)\end{tabular} & \begin{tabular}[c]{@{}c@{}}$E_J/h$\\ (GHz)\end{tabular} & \begin{tabular}[c]{@{}c@{}}$E_L/h$\\ (GHz)\end{tabular} & \begin{tabular}[c]{@{}c@{}}$T_{\text{eff}}$\\ (mK)\end{tabular} \\ \hline
D1\_1 & 0.5:0.5 & 1000$^{\circ}$C & 10 & 9.12 & 0.05 & 3.57 & 1.0 & 280.0 & 4.7(0.7) & 1601(755) & 1.20(0.07) & 1.42 & 4.07 & 0.39 & 23 \\
D1\_2 & 0.5:0.5 & 1000$^{\circ}$C & 10 & 9.12 & 0.05 & 3.57 & 1.0 & 320.0 & 5.5(2.5) & 3104(2008) & 1.25(0.11) & 1.37 & 4.42 & 0.35 & 14 \\
D1\_3 & 0.5:0.5 & 1000$^{\circ}$C & 10 & 9.12 & 0.05 & 3.57 & 1.0 & 340.0 & 12.9(4.3) & 2458(1088) & 1.23(0.06) & 1.30 & 5.25 & 0.32 & 15 \\
D2\_1 & 0.5:0.5 & 1000$^{\circ}$C & 10 & 8.82 & 0.08 & 2.84 & 1.0 & 226.9 & 9.5(1.3) & 403(204) & 1.07(0.08) & 1.39 & 5.77 & 0.47 & 20 \\
D2\_2 & 0.5:0.5 & 1000$^{\circ}$C & 10 & 8.82 & 0.08 & 2.84 & 2.5 & 567.1 & 10.3(2.2) & 1544(835) & 1.22(0.08) & 1.30 & 5.53 & 0.50 & 20 \\
D3\_1 & 0.7:0.3 & 1000$^{\circ}$C & 10 & 9.8 & 0.43 & 1.01 & 0.2 & 398.9 & 2.8(0.5) & 30(9) & 0.97(0.05) & 1.52 & 3.88 & 0.89 & 22 \\
D4\_1 & 0.7:0.3 & 900$^{\circ}$C & 10 & 9.65 & 0.47 & 0.98 & 0.55 & 1119.0 & 8.9(1.3) & 132(49) & 1.07(0.05) & 1.34 & 5.91 & 0.50 & 57 \\
D4\_2 & 0.7:0.3 & 900$^{\circ}$C & 10 & 9.65 & 0.47 & 0.98 & 0.65 & 1322.5 & 6.1(1.2) & 46(16) & 0.94(0.05) & 1.33 & 5.83 & 0.53 & 14 \\
D5\_1 & TiN & N/A & 10 & 10.7 & 0.99 & 0.58 & 0.2 & 1000.0 & 3.6(0.8) & 6(1) & 0.74(0.02) & 1.40 & 3.77 & 0.65 & 15 \\
D5\_2 & TiN & N/A & 10 & 10.7 & 0.99 & 0.58 & 0.25 & 1250.0 & 3.2(0.7) & 9(2) & 0.80(0.02) & 1.36 & 4.33 & 0.68 & 14 \\
D6\_1 & TiN & N/A & 10 & 11.1 & 1.07 & 0.53 & 0.2 & 1000.0 & 3.8(1.1) & 20(4) & 0.87(0.03) & 1.39 & 4.74 & 0.60 & 30 \\
D6\_2 & TiN & N/A & 10 & 11.1 & 1.07 & 0.53 & 0.25 & 1250.0 & 2.2(0.7) & 20(5) & 0.88(0.03) & 1.38 & 4.57 & 0.64 & 31 \\
D7\_1 & 0.5:0.5 & 900$^{\circ}$C & 20 & 8.66 & 0.27 & 1.44 & 0.5 & 739.4 & 2.9(0.8) & 58(20) & 0.94(0.05) & 1.36 & 3.96 & 0.66 & 31 \\
D7\_2 & 0.5:0.5 & 900$^{\circ}$C & 20 & 8.66 & 0.27 & 1.44 & 0.55 & 813.3 & 2.7(0.8) & 47(11) & 0.92(0.03) & 1.36 & 3.81 & 0.68 & 26 \\
D7\_3 & 0.5:0.5 & 900$^{\circ}$C & 20 & 8.66 & 0.27 & 1.44 & 0.6 & 887.3 & 2.7(0.6) & 17(3) & 0.84(0.02) & 1.38 & 3.01 & 0.73 & 23 \\
D7\_4 & 0.5:0.5 & 900$^{\circ}$C & 20 & 8.66 & 0.27 & 1.44 & 0.6 & 887.3 & 3.3(0.7) & 20(3) & 0.83(0.02) & 1.29 & 2.74 & 0.56 & 14 \\
D8\_1 & 0.5:0.5 & 900$^{\circ}$C & 20 & 8.87 & 0.25 & 1.46 & 0.55 & 934.5 & 4.3(1.4) & 111(32) & 0.96(0.04) & 1.37 & 6.03 & 0.59 & 22 \\
D8\_2 & 0.5:0.5 & 900$^{\circ}$C & 20 & 8.87 & 0.25 & 1.46 & 0.65 & 1104.4 & 3.6(1.1) & 104(30) & 0.95(0.04) & 1.38 & 5.80 & 0.57 & 28 \\
D9\_1 & 0.5:0.5 & 900$^{\circ}$C $\mathrm{N}_2$ & 20 & 8.94 & 0.16 & 1.81 & 0.55 & 591.7 & 2.7(0.9) & 160(35) & 1.01(0.03) & 1.43 & 4.72 & 0.61 & 22 \\
D9\_2 & 0.5:0.5 & 900$^{\circ}$C $\mathrm{N}_2$ & 20 & 8.94 & 0.16 & 1.81 & 0.65 & 699.3 & 3.0(0.8) & 123(29) & 0.99(0.03) & 1.42 & 4.69 & 0.63 & 36 \\
D10\_1 & 0.5:0.5 & 1000$^{\circ}$C & 20 & 8.95 & 0.22 & 1.55 & 0.55 & 789.4 & 1.7(0.8) & 17(4) & 0.79(0.03) & 1.39 & 4.92 & 0.83 & 12 \\
D10\_2 & 0.5:0.5 & 1000$^{\circ}$C & 20 & 8.95 & 0.22 & 1.55 & 0.65 & 932.9 & 2.3(0.8) & 19(4) & 0.81(0.03) & 1.39 & 4.40 & 0.85 & 18 \\
D11\_1 & 0.5:0.5 & 1100$^{\circ}$C & 20 & 9.25 & 0.43 & 1.06 & 0.45 & 1386.4 & 3.3(0.7) & 18(3) & 0.83(0.02) & 1.37 & 5.42 & 0.55 & 35 \\
D11\_2 & 0.5:0.5 & 1100$^{\circ}$C & 20 & 9.25 & 0.43 & 1.06 & 0.45 & 1386.4 & 5.1(1.2) & 18(4) & 0.82(0.03) & 1.38 & 5.57 & 0.53 & 12 \\
D12\_1 & 0.6:0.4 & 1000$^{\circ}$C & 20 & 10.1 & 0.57 & 0.85 & 0.2 & 1000.0 & 6.0(0.6) & 20(5) & 0.99(0.04) & 1.36 & 5.53 & 0.84 & 18 \\
D12\_2 & 0.6:0.4 & 1000$^{\circ}$C & 20 & 10.1 & 0.57 & 0.85 & 0.25 & 1250.0 & 1.8(0.3) & 13(3) & 0.92(0.04) & 1.39 & 4.10 & 0.85 & 19 \\
D12\_3 & 0.6:0.4 & 1000$^{\circ}$C & 20 & 10.1 & 0.57 & 0.85 & 0.2 & 1000.0 & 2.9(0.8) & 16(4) & 0.91(0.03) & 1.38 & 4.98 & 0.82 & 22 \\
D12\_4 & 0.6:0.4 & 1000$^{\circ}$C & 20 & 10.1 & 0.57 & 0.85 & 0.25 & 1250.0 & 2.4(0.7) & 15(3) & 0.90(0.03) & 1.36 & 5.27 & 0.81 & 21 \\
D13\_1 & 0.6:0.4 & 1000$^{\circ}$C & 20 & 9.92 & 0.5 & 0.92 & 0.2 & 1000.0 & 2.3(0.6) & 37(12) & 1.00(0.05) & 1.39 & 4.61 & 0.65 & 20 \\
 \hline\hline
\end{tabularx}
\end{table*}

\begin{figure} [t]
      \centering
      \includegraphics[width=8.6cm]{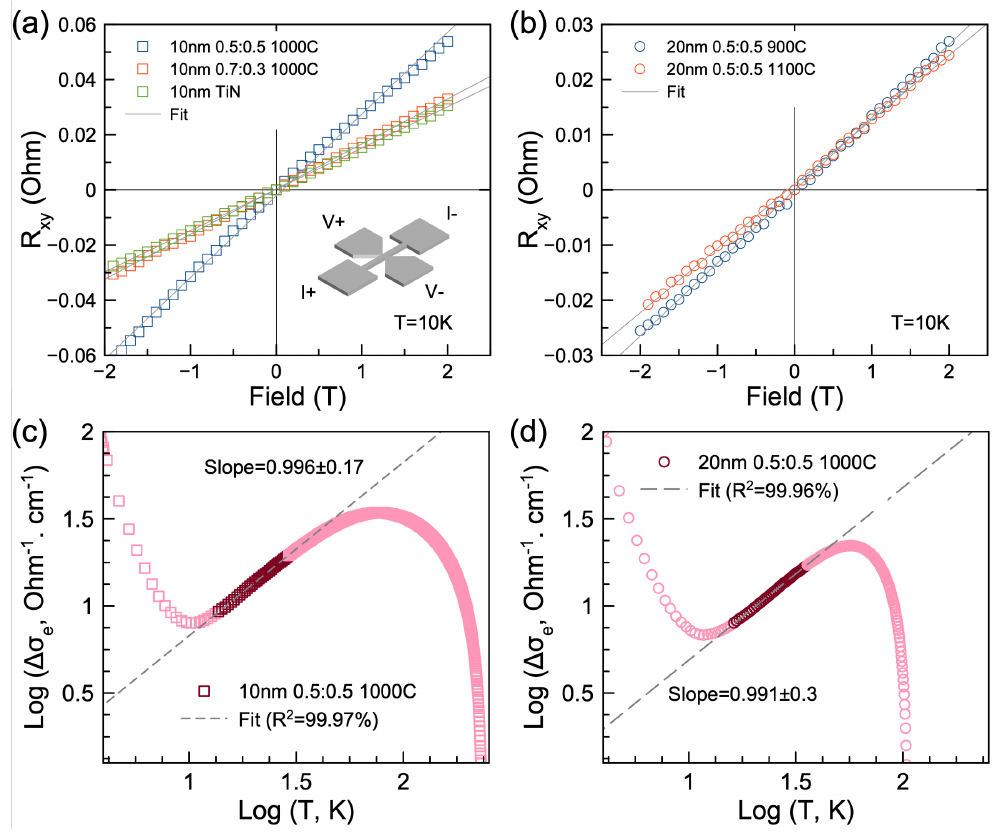}
      \caption{(a-b) The transverse resistance $R_{xy}$ as a function of applied field $H$ for a range of exemplary samples measured at 10~K. Slight deviation from a linear fitting was only observed in the 10 nm-thick Ti\textsubscript{0.5}Al\textsubscript{0.5}N films at the strong-disorder limit. (c-d) Log-log plots of $\Delta\sigma_e$ as a function of the temperature $T$ for two films. The $d\sigma_e/dT>0$ part of all data can be fitted linearly with a slope $\approx 1$. At lower temperatures, the $d\sigma_e/dT<0$ is due to paraconductivity, while at higher temperatures, thermally activated electron-phonon scattering starts to dominate.}
      \label{fig:TANNQFigS11} 
\end{figure}

Lastly, we discuss the fundamental transport properties of the film set and understand the implications. The standard transverse resistance ($R_{xy}$) measurements as a function of the vertically applied fields were used to extract the carrier concentration, and a number of exemplary samples are shown (\autoref{fig:TANNQFigS11}a-b, where the inset illustrates the electrical leads of the setup). The measurements were all conducted at 10~K with a supplied DC current of 1~$\mu$A, and the carrier densities were calculated using $1/(en_et)=R_{xy}/(\mu_0H)$, where $e$ is the electrical charge, $n_e$ is the charge carrier density at 10~K used for further analysis, and $H$ is the amplitude of the external fields. For all the tested samples, the carrier density ranges from $\sim 2.2\times10^{22}$ to $\sim 4.2\times10^{22}$~cm\textsuperscript{3}. Such high carrier concentration has confirmed the metallic nature of the films, and correspondingly the applicability of the free-electron approximation. 

The electron-transport properties were understood by analyzing the temperature dependent part of the conductivity. Here, the temperature dependent part is given by $\Delta\sigma_e=\sigma_e-\sigma_e(0)$, where $\sigma_e$ is the electrical conductivity and $\sigma_e(0)$ is the conductivity at zero temperature. Since the sample is superconducting at low temperature, $\sigma_e(0)$ was obtained from either linear extrapolation of the conductivity above $T_c$ or more directly from $R_{xx}$ measurements under a large applied field ($>$6~Tesla). We found that both approaches give similar results, and the choice has minor impacts on the conclusions. By plotting the $\Delta\sigma_e$ vs $T$ in a log-log scale, the $d\sigma_e/dT>0$ regime from $\sim$15 K to $\sim$30 K can be fitted for all samples, irrespective of the different material preparation methods (\autoref{fig:TANNQFigS11}c-d). The fittings typically yield a slope of $0.99\approx1$ with fitting fidelity $R^2 > 99.96\%$. As our system does not have strong spin-orbit coupling, the $d\sigma_e/dT>0$ behaviors in the metallic samples are typically mediated by either the weak localization or the electron-electron interaction effects~\cite{Altshuler1979, Lee1985, Kaveh1981, Kaveh1981a, Kaveh1982}. In particular, the temperature dependent part of the conductivity as a result of the two effects can be written as
\begin{equation}
    \label{eqn:eqn1}
    \Delta\sigma_e=\frac{e^2}{\pi^2\hbar}\Bigr[\frac{1}{L_{IN}}+f\cdot\frac{1}{L_{INT}}\Bigl],
\end{equation}
where $L_{IN}=(1/2ll_{IN})^{1/2}$ is the length scale of the constructive interference given the elastic mean free path $l$ and the inelastic mean free path $l_{IN}$. $L_{INT}$ is the interaction length scale of the electron-electron interaction that give rise to the energy-level broadening, and $f$ is a factor that could be either positive or negative, depending on the nature of the system and the temperature range. In the 3D limit we have
\begin{equation}
    \label{eqn:eqn2}
    L_{IN}^{-1}\propto T^{-P/2},
\end{equation}
\begin{equation}
    \label{eqn:eqn3}
    L_{INT}^{-1}\propto (T/D)^{1/2} \propto T^{1/2},
\end{equation}
where the factor $P$ is determined by the specific scattering mechanism and $D$ is the electron diffusivity. While in the 2D limit we have
\begin{equation}
    \label{eqn:eqn4}
    L_{IN}^{-1}\propto \ln{T},
\end{equation}
\begin{equation}
    \label{eqn:eqn5}
    L_{INT}^{-1}\propto \ln{k_BT\tau/\hbar},
\end{equation}

Thus, assuming an electron-phonon mediated scattering process ($P=2$), the films measured in this studies ubiquitously showed weak-localization features in the 3D limit. This is consistent with the fact that the measured mean free paths $l$ of the samples are on the order of angstroms, much smaller than the film thicknesses (see Table~\ref{tab:table1}). We also note that at temperatures higher than 30 K~but below 40~K, $\Delta\sigma$ can be fitted by $\ln{T}$ with $R^2 \sim 99.6\%$. We do not take this as the major physical process in the samples due to the fact that the electron-electron interaction typically happens at much lower temperatures, where the fitting range and fidelity are much worse.

It is worth noting that we take disorder as the sole indicator of the material properties, regardless of the different experimental techniques applied to tune the microstructures. The rationalization and experimental grounding of this approach are as the following. First, materials with distinct microstructures but similar disorder ($\rho_{xx}$) are found to be comparable in terms of qubit coherence. For instance, qubits with 10 nm-thick Ti\textsubscript{0.7}Al\textsubscript{0.3}N films annealed at 900$^{\circ}$C have roughly the same noise amplitude and dielectric loss tangent as those with 20 nm-thick Ti\textsubscript{0.5}Al\textsubscript{0.5}N films annealed at 1100$^{\circ}$C (see Table~\ref{tab:table1} and the main text). In other words, even though the impact of the different microstructures is yet clear, the dominant tuning knob of the qubit coherence is suggested to be the disorder. Second, as discussed above, for all the tested Ti\textsubscript{x}Al\textsubscript{1-x}N films, the transport behaviors revealed a $\Delta \sigma_e \propto T$ relationship in a similar temperature window from $\sim$15 K to $\sim$30 K, indicative of weak localization in the 3D regime. It is thus a reasonable postulation that the properties of the tested films, irrespective of the material preparation methods, share a common scattering mechanism but only differ in the total scattering-center density.

\begin{figure}[b]
      \centering
      \includegraphics[width=8.6cm]{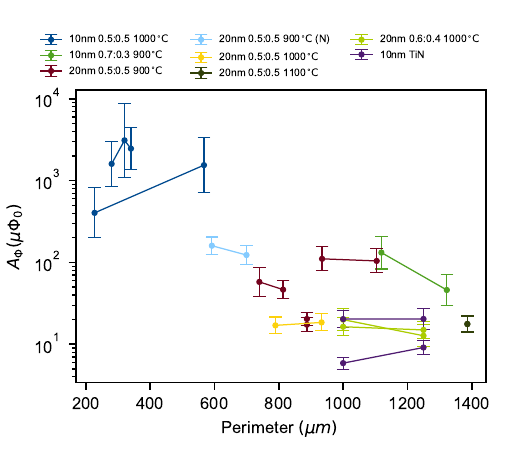}
      \caption{The $1/f^\alpha$ flux noise amplitude $\ATtwo$ plotted against the perimeter of the devices. Connected solid dots are devices from the same chip that have a fixed aspect ratio $p/w$, where the perimeter and width are simultaneously varying. }
      \label{fig:TANNQFigS12} 
\end{figure}

\section{Geometric dependency of the flux noise amplitudes}

Here, we show the dependency of the qubit coherence on the wire geometries. In terms of the wire geometries, since we find that the dielectric loss is mostly unaffected by these variations, we solely focus on analyzing the flux noise amplitudes in relation to the inductor-wire geometries (\autoref{fig:TANNQFigS12}). It can be observed from the plot that, within the margin of error, the noise amplitude $\ATtwo$ remains constant for a specific aspect ratio $p/w$. Although we did not extensively vary the perimeter as limited in the scope of this work, we contend that the flux noise amplitude of our devices is primarily influenced by the phenomenological spin defect density $\sigma$ and the aspect ratio of the wire. This finding aligns with the implications of the fluctuating spin model~\cite{Bialczak2007, Koch2007, Anton2013, Braumuller2020} for $\lambda > w \gg t$.

\clearpage
\bibliography{TAN-NQ-Ref_Supp}